\documentclass[11pt]{article}
\usepackage{graphicx,dcolumn,amsmath,latexsym,amssymb,natbib}
\usepackage{amsfonts,subfigure,caption}
\usepackage{color,bm}
\usepackage{placeins}
\usepackage{hyperref}

\newtheorem{thm}{Theorem}[section]

\newtheorem{remark}[thm]{Remark}

\newtheorem{problem}[thm]{Problem}
\numberwithin{equation}{section}

\newcommand{\IGNORE}[1]{}
\newcommand{\ignore}[1]{}

\newcommand{\im}{\operatorname{Im}}

\newcommand{\mbb}[1]{\mathbb{#1}}

\newcommand{\mc}[1]{\mathcal{#1}}

\newcommand{\la}{\langle}
\newcommand{\ra}{\rangle}

\newcommand{\n}{\mathbf{\tilde{n}}}
\newcommand{\PP}{\operatorname{P}}
\newcommand{\PR}{\operatorname{R}}
\DeclareMathOperator*{\argmax}{argmax}
\newcommand{\bds}{\boldsymbol}
\newcommand{\emach}{\epsilon_{\text{machine}}}
\newcommand{\tol}{\text{tol}}

\newcommand{\TG}{{\text{TG}}}
\newcommand{\rand}{{\text{rand}}}

\newcommand{\Kerr}{{\text{K}}}

\newcommand{\Hou}{{\text{H}}}
\newcommand{\Trel}{{T_{\text{rel}}}}

\def\bnabla{\boldsymbol{\nabla}}
\def\u{\boldsymbol{u}}
\def\vv{\boldsymbol{v}}
\def\w{\boldsymbol{w}}
\def\d{\boldsymbol{d}}
\def\j{\boldsymbol{j}}
\def\bk{\boldsymbol{k}}
\def\n{\boldsymbol{n}}
\def\x{\boldsymbol{x}}

\def\bbeta{\boldsymbol{\eta}}
\def\bxi{\boldsymbol{\xi}}
\def\bomega{\boldsymbol{\omega}}
\def\RR{\mathbb{R}}
\def\TT{\mathbb{T}}
\def\CC{\mathbb{C}}

\def\tPhi{\widetilde{\Phi}}
\def\teta{\widetilde{\boldsymbol{\eta}}}

\def\M{\mathcal{M}}
\def\T{\mathcal{T}}

\begin{document}

\title{\vspace*{-2.5cm} Systematic search for singularities in 3D Euler flows }
\author{Xinyu Zhao$^1$ and Bartosz Protas$^{1,}$\thanks{Corresponding Author, Email: {\tt bprotas@mcmaster.ca}}} \date{
  $^1$Department of Mathematics and Statistics, McMaster University \\
  Hamilton, ON, Canada \\ \medskip \today}
\maketitle

\begin{abstract}
  We consider the question whether starting from a smooth initial
  condition 3D inviscid Euler flows on a periodic domain 
  may develop singularities in {a} finite time. Our point of departure
  is the well-known result by \citet{Kato1972}, which asserts the
  local existence of classical solutions to the Euler system in the
  Sobolev space $H^m$ for $m > 5/2$. Thus, the potential formation of
  a singularity must be accompanied by an unbounded growth of the
  $H^m$ norm of the velocity field as the singularity time is
  approached. We perform a systematic search for ``extreme'' Euler
  flows that may realize such a scenario by formulating and solving a
  PDE-constrained optimization problem where the $H^3$ norm of the
  solution at a certain fixed time $T > 0$ is maximized with respect
  to the initial data subject to suitable normalization constraints.
  This problem is solved using a state-of-the-art Riemannian conjugate
  gradient method where the gradient is obtained from solutions of an
  adjoint system.  Computations performed with increasing numerical
  resolutions demonstrate that, as asserted by the theorem of
  \citet{Kato1972}, when the optimization time window $[0, T]$ is
  sufficiently short, the $H^3$ norm remains bounded in the extreme
  flows found by solving the optimization problem, which indicates
  that the Euler system is well-posed on this ``short'' time interval.
  On the other hand, when the window $[0, T]$ is long, possibly longer
  than the time of the local existence asserted by Kato's theorem,
  then the $H^3$ norm of the extreme flows diverges upon resolution
  refinement, which indicates a possible singularity formulation on
  this ``long'' time interval.  The extreme flow obtained on the long
  time window has the form of two colliding vortex rings and is
  characterized by certain symmetries. In particular, the region of
  the flow in which a singularity might occur is nearly axisymmetric.
    
  \end{abstract}




\section{Introduction}

The motion of an ideal (inviscid) incompressible fluid on a domain
$\Omega \subseteq \mbb R^3$ is described by the Euler equations
\begin{equation}\label{eq:Euler}
\begin{aligned}
 \frac{\partial \bds u}{\partial t} + (\bds u \cdot \bnabla)\bds u &= - \nabla p, &\qquad &(\bds x, t) \in  \Omega \times (0, T],\\
\bnabla \cdot \bds u &= 0, &\qquad &(\bds x, t) \in \Omega \times (0, T], \\
\bds u|_{t = 0} &= \bbeta,  &\qquad  &\bds x\in\Omega,
 \end{aligned}
\end{equation}
where $\bnabla = [\partial_1, \partial_2, \partial_3]^T$ is the
gradient operator, $\bds u = \u(\x,t) = [u_1, u_2, u_3]^T$ is the
velocity field, $p = p (\bds x, t)$ is the scalar pressure and $T > 0$
is the length of the time window considered. We assume the flow domain
to be a periodic unit cube $\Omega = \mbb T^3: = \mbb R^3/ \mbb Z^3$,
where ``:='' means ``equal to by definition''.  In system
\eqref{eq:Euler}, the symbol $\bbeta$ denotes the initial
condition for the velocity which is assumed to satisfy
$\bnabla \cdot \bbeta = 0$. We will use the notation $\u(t; \bbeta)$
to indicate the dependence of the solution of system \eqref{eq:Euler}
at time $t$ on $\bbeta$.  We are interested in the possibility of a
spontaneous formation of finite-time singularities in solutions of the
Euler system \eqref{eq:Euler} equipped with smooth initial data, 
the so-called ``blow-up problem''.  It remains one of the
central open questions in mathematical fluid mechanics \citep{gbk08}
and is closely related to the corresponding regularity problem for the
viscous Navier-Stokes system \citep{d09,Robinson2020}, which has been
recognized by the Clay Mathematics Institute as one of its
``millennium problems'' \citep{f00}.  Our goal in this paper is to
undertake a systematic search for possible singularities in the Euler
system \eqref{eq:Euler} using methods of numerical PDE optimization.

The first results asserting local existence of classical solutions to
the Euler system were obtained by \citet{Lichtenstein1925} for
H\"older-regular initial data $\bbeta \in C^{1,\alpha}(\RR^3)$ with $0
< \alpha < 1$. The local existence of classical solutions in Sobolev
spaces was then established by \citet{Kato1972} and is summarized in
the following theorem
\begin{thm} \label{thm:Hm} If $\bbeta \in H^m (\mbb T^3)$ for some $m
  > 5/2$ and satisfies $\bnabla \cdot \bbeta = 0$, then there exists a
  time $T = T\left(\| \bbeta \|_{H^m}\right)>0$ such that
  \eqref{eq:Euler} has a unique solution $\bds u(\cdot;\bbeta) \in
  C([0, T]; H^m)$ $\bigcap C^1([0,T]; H^{m-1})$.
\end{thm}
\noindent
The study of the local well-posedness of the Euler system
\eqref{eq:Euler} with analytic initial data began with the work of
\citet{Bardos1977analyticity}.

Another well-known conditional regularity result is the
Beale-Kato-Majda (BKM) criterion \citep{bkm84, Chen:2012:BKM} which states that a
smooth solution $\bds u$ of the Euler system develops a singularity at
$t = T^*$ if and only if
\begin{equation}\label{eq:BKM:vort}
{\lim_{t \rightarrow T^*} \int_{0}^{t} ||\bds \omega(\tau)||_{L^\infty} \;d\tau = \infty,}
\end{equation}
where $\bds \omega := \nabla \times \bds u$ is the vorticity field of
the flow. The sufficiency of this condition can be deduced from
Theorem \ref{thm:Hm} using a Sobolev inequality \citep{af05}
\begin{equation}
{\int_{0}^{t} ||\bds \omega(\tau)||_{L^\infty} \;d\tau \leq t \sup_{0\leq \tau\leq t} ||\nabla \bds u(\tau)||_{L^\infty}
\leq C t \sup_{0\leq \tau \leq t} ||\bds u(\tau)||_{H^m}, \quad m > 5/2,}
\end{equation}
whereas its necessity is a result of the inequality \citep{mb02}
\begin{equation}\label{eq:BKM:necessary}
||\bds u(t)||_{H^m} \leq ||\bbeta||_{H^m}\exp\left\{C_1 \exp\left(C_2\int_{0}^{{t}} ||\bds \omega(\tau)||_{L^\infty} \;d\tau \right)\right\}.
\end{equation}
The double exponential on the right hand side (RHS) of
\eqref{eq:BKM:necessary} suggests that, should blow-up indeed occur at
some $t = T^*$, we can expect a much more rapid growth of
$\| \u(t) \|_{H^m}$, $m > 5/2$, than that of $\| \bomega(t)
\|_{L^\infty}$, as $t \rightarrow T^*$. There have been various
refinements of the BKM criterion, where the $L^\infty$ norm of
vorticity in \eqref{eq:BKM:vort} is replaced with a norm in the BMO
space \citep{Kozono2000BMO} or a Besov space \citep{Chae2001Besov}. In
addition, there are also geometric criteria for blow-up \citep{CFM:96,
  Hou:05:Geometric}, in which the direction of the vorticity $\bds
\omega/|\bds \omega|$ plays a crucial role.  As was shown in
\cite{Gibbon2013}, the relative ordering of suitably rescaled
vorticity moments provides information about the degree of depletion
of the nonlinearity, and hence also about the regularity of solutions.
On the other hand, the regularity of weak solutions of the Euler
system is related to Onsager's conjecture concerning energy
dissipation in such flows \citep{ConstantinETiti1994} and significant
progress has been made recently as regards the non-uniqueness of weak
solutions with low regularity \citep{DeLSze19}.

With respect to blow-up scenarios, \citet{Elgindi:21:blowup} proved
that there exist swirl-free axisymmetric solutions of the Euler system
corresponding to initial data in $C^{1,\alpha}(\RR^3)$ that can
develop finite-time singularities.  \citet{ElgindiJeong2018} showed
the existence of finite-time singularities in strong solutions to
axisymmetric Euler equations on exterior domains with
``hourglass''-shaped boundaries.  Most recently,
\citet{Hou:2022:proof} proved a nearly self-similar blow-up of
solutions to the 3D axisymmetric Euler equations with smooth initial
data on cylindrical domains with solid boundaries. We emphasize that
in the last two cases the presence of the solid boundary is key to the
formation of the singularity.  

The computational studies exploring the possibility of finite-time
blow-up in the Euler system include
\cite{bmonmu83,ps90,b91,k93,p01,bk08,oc08,o08,ghdg08,
  gbk08,h09,opc12,bb12,k13,opmc14,CampolinaMailybaev2018,Larios2018},
all of which considered problems subject to periodic boundary
conditions in all three spatial directions.  We also mention the
studies by \cite{mbf08} and \cite{sc09}, along with references found
therein, in which various complexified forms of the Euler equations
were investigated. The idea of this approach is that, since solutions
to complexified equations have singularities in the complex plane,
singularity formation in the real-valued problem is manifested by the
collapse of the complex-plane singularities onto the real axis.  Some
of the investigations \citep{bb12,opc12} hinted at the possibility of
singularity formation in a finite time. In this connection we also
highlight the {computational} investigations of \citet{lh14a,lh14b} in
which blow-up was {documented} in axisymmetric Euler flows on a
bounded cylindrical domain. This mechanism of singularity formation
involves an interaction with the solid boundary and was recently
validated with rigorous mathematical analysis by
\citet{Hou:2022:proof}. We also mention an investigation by
\citet{Hou:22:Euler} who provided evidence for blow-up in axisymmetric
Euler flows on bounded domains in which singularity occurs away from
the boundaries. In contrast to most other studies, the works of
\citet{lh14a,lh14b,Hou:22:Euler} relied on adaptive mesh refinement
employed to resolve fine structures in flows at the edge of
regularity. In this context, we also mention the investigation by
\citet{Yin:2021:CM:3D} where the Euler system \eqref{eq:Euler} was
solved in the Lagrangian setting using the characteristic mapping
method.  Regarding weak solutions and Onsager's related conjecture,
\citet{Fehn2022} recently presented numerical evidence for
nonvanishing energy dissipation in weak solutions of the Euler system
\eqref{eq:Euler} which was obtained using a method based on a
discontinuous Galerkin approximation.

While in the aforementioned computational studies the initial
conditions for the Euler system were chosen in an ad-hoc manner,
albeit one usually motivated by deep physical considerations, here we
follow a fundamentally different approach wherein the initial
condition leading to the most singular, in a mathematically precise
sense, solutions is sought systematically by solving a suitably
constrained PDE optimization problem.  This approach was originally
proposed by \citet{ld08} as a way to determine incompressible velocity
fields maximizing the instantaneous growth rate of enstrophy in 3D
Navier-Stokes flows.  It was later extended by \citet{ap11a, ap13a,
  ap16}, \citet{Yun2018} to study extreme behavior in
solutions of different hydrodynamic models such as variants of the
Burgers equation.  Recently, \citet{KangYunProtas2020} and
\citet{KangProtas2021} applied this method to search for potential
singularities in 3D Navier-Stokes flows based on some classical
conditional regularity results. Highlights of this more than
decade-long research program are summarized in the review paper by
\citet{p21a}. The present investigation represents a first application
of this framework to an inviscid problem.

Guided by the local well-posedness result in Theorem \ref{thm:Hm}, we
aim to find an initial condition $\bbeta\in H^m$ subject to certain
constraints, such that the $H^m$ norm of the corresponding
solution of the Euler system \eqref{eq:Euler} is maximized at a
prescribed time $T$.  The desired initial conditions are thus found as
local maximizers of a constrained PDE optimization problem with the
square of the $H^m$ (semi)norm used as the objective functional where
for concreteness we set $m = 3$.  We solve this optimization problem
for different time intervals $[0, T]$ using a Riemannian conjugate
gradient method \citep{ams08}, where the gradient is conveniently
computed from solutions of a suitably-defined adjoint system. Both the
Euler system \eqref{eq:Euler} and the adjoint system as well as 
different diagnostic quantities are approximated numerically using
pseudospectral methods.

In analogy to the work of \citet{Fehn2022, Guo:2022:blowup}, we adopt
an indirect approach to distinguish between regular and singular
evolution based on resolution refinement.  When solving the
optimization problem on a ``short'' time interval $[0, T]$, if the
objective functional approximated using different resolutions
converges to a finite value as the resolution is refined, then we
conclude the Euler system \eqref{eq:Euler} is well-posed on this short
interval.  However, when the interval $[0, T]$ is ``long'', presumably
longer than the minimum time of existence guaranteed by
Theorem~\ref{thm:Hm}, the objective functional evaluated at the
optimal solutions may diverge upon resolution refinements.  {Here
  ``short'' and ``long'' times are defined in relation to the interval
  of local existence guaranteed by Theorem~\ref{thm:Hm}.  More
  specifically, a ``short'' time is assumed to be within that
  interval, whereas a ``long'' time is outside. Theorem~\ref{thm:Hm}
  does not provide a precise numerical value for the interval of local
  existence and in} practice, based on numerical experiments, we use
$T = 25$ and $T = 75$ for the short and long optimization time
windows.  We do in fact observe these two distinct behaviors and
conclude that in the latter scenario the divergence of the objective
functional on the long time interval $[0,75]$ may signal a potential
singularity formation in the flow. Although a definitive conclusion
cannot be drawn due to numerical limitations, the behavior of
different diagnostic quantities does not contradict the possibility of
a singularity formation.  The corresponding flow features two
colliding vortex rings and while these flow structures are quite
deformed, the flow does exhibit certain nontrivial symmetries.  In
particular, the region in the flow where the potential singularity may
occur is nearly axisymmetric.
 
The structure of the paper is as follows: in Section \ref{sec:optim} we
formulate an optimization problem designed to elucidate the most
extreme behavior possible in an Euler flow and in
Section \ref{sec:approach} we describe a Riemannian gradient-based
approach we use to solve this problem together with its numerical
discretization; the extreme flows found by the optimization algorithm
on ``short'' and ``long'' time intervals are discussed in
Section \ref{sec:results}; finally, conclusions and outlook are deferred
to Section \ref{sec:final}.

\section{Optimization problem}
\label{sec:optim}

Before introducing the optimization problem, let us first define the
function space in which its solutions will be sought. We begin by
defining the norm in the Sobolev space $H^m(\TT^3)$ as
\citep{af05}
\begin{equation}
\|  \bds u \|_{H^m}  :=  \left[ \sum_{\bds j \in \mathbb Z^3} 
\left(1+|2\pi\bds j|^2\right)^m   |\hat{\bds u}_{\bds j}|^2 \right]^{1/2}, \qquad m \in \RR,
\label{eq:Hm}
\end{equation}
where $\hat{\bds u}_{\bds j}$ is the Fourier coefficient of $\u$
corresponding to the wavevector $\bds k := 2\pi \j = [2\pi j_1, 2\pi
j_2, 2\pi j_3]$. The corresponding homogeneous seminorm, denoted
$\|\cdot\|_{\dot{H}^m}$, is obtained from \eqref{eq:Hm} by dropping
the constant ($\j$-independent) terms. For simplicity, we
  will hereafter refer to it as the $H^m$ seminorm.

\subsection{Functional setting} \label{sec:function}

Since the Euler system \eqref{eq:Euler} is locally well posed in $H^m(\TT^3)$,
$m>5/2$, cf.~Theorem \ref{thm:Hm}, it may appear natural to look for
optimal initial data $\bbeta$ in that space. However, we are
interested in finite-time singularities potentially arising in
classical solutions, whereas initial conditions constructed in such
spaces will in general not be smooth or real-analytic.  Moreover,
solving the Euler system \eqref{eq:Euler} with such initial data would
not allow us to benefit from the exponential convergence of the
pseudospectral methods used in this study,
cf.~Section \ref{sec:num:discretization}. We will thus consider an
extended Gevrey space $G^\sigma$ with $\sigma > 0$ of real-analytic
functions defined on $\TT^3$ and will endow it with the inner product
\begin{equation}\label{eq:Gevrey:def}
\begin{aligned}
\forall \bds v, \bds u\in G^{\sigma}, \quad 
\left\la\bds v, \bds u \right\ra_{G^{\sigma}} 
:= &\sum_{\bds j \in \mathbb Z^3} 
(1+|2\pi\bds j|^2)^me^{4\pi\sigma|\bds j|} \hat{\bds v}_{\bds j} \cdot \overline{\hat{\bds u}}_{\bds j}\\
=& \int_{\mbb T^3}(1+|D|^2)^me^{2\sigma|D|} \bds v \cdot \bds u \, d\bds x,
\end{aligned}
\end{equation}
where overbar denotes complex conjugation, whereas the operators $|D|$
and $e^{\sigma|D|}$ are defined via
\begin{equation}
\left[\widehat{|D| \bds v}\right]_{\bds j} := 2\pi|\bds j| \hat{\bds v}_{\bds j}, \qquad \qquad 
\left[\widehat{e^{\sigma |D|}\bds v}\right]_{\bds j} := e^{2\pi\sigma|\bds j|}\hat{\bds v}_{\bds j}.
\label{eq:D}
\end{equation}
In the setting of our problem, the Gevrey space can be regarded a
linear subspace of the Sobolev space $H^m(\TT^3)$, i.e.,
\begin{equation}
G^\sigma := \left\{ \vv \in H^m(\TT^3) \ : \ \| \vv \|_{G^\sigma} = \left\langle \vv,\vv \right\rangle^{1/2}_{G^\sigma} < \infty \right\}, \quad m > \frac{5}{2}, \quad \sigma > 0.
\label{eq:G}
\end{equation}

\begin{remark}
  It follows from definitions \eqref{eq:Gevrey:def} and \eqref{eq:G}
  that the Gevrey spaces have the property
\begin{equation}
G^{\sigma_2} \subset G^{\sigma_1} \subset G^{0} = H^m, 
\qquad 0 < \sigma_1 < \sigma_2.
\end{equation}
Our strategy for choosing the value of $\sigma$ will be discussed in Section \ref{sec:Tshort}.
\end{remark}

Since the initial condition $\bbeta$ in system \eqref{eq:Euler} needs to
be divergence-free and the quantity $\int_{\TT} \bbeta \, d\x$ is an
invariant of motion, we introduce the following subspace in which
optimal initial conditions will be sought
\begin{equation}\label{eq:space:initial:condition}
V := \{ \bds v  \in G^{\sigma}: \,
\bnabla \cdot \bds v = 0, \quad \int_{\mbb T^3} \bds v \, d \bds x= \bds 0\}.
\end{equation}

The Gevrey regularity is closely related to the analyticity of complex
extensions of functions in $G^\sigma$, i.e., $\bds u(\bds z):= \u(\bds
x + i\bds y)$, where $i := \sqrt{-1}$. If $\bds u\in G^\sigma$ can be
extended to a strip $S_{\delta} := \{\bds z = [z_1, z_2,
  z_3]^T \in \mbb C^3: |\im(\bds z_1)| + |\im(\bds z_2)| + |\im(\bds
  z_3)| < \delta \}$, so that $\bds u(\bds z)$ does not have any
  singularities in $S_\delta$, then the largest such value of $\delta$
  is called the width of the analyticity strip and we have $0 <
\sigma \le \delta$.  Furthermore, if $\u = \u(t)$ is a smooth
time-dependent solution of the Euler system \eqref{eq:Euler}, then the
width of the analyticity strip is a function of time, $\delta =
\delta(t)$.

As regards the analyticity of solutions of the Euler system
\eqref{eq:Euler}, among related works we mention the study by
\citet{Bardos1977analyticity} which showed that if the initial data
$\bbeta$ is analytic, the solution $\u(t; \bbeta)$ also remains
analytic as long as it is well defined in a certain H\"older space.
The relation between the BKM condition \eqref{eq:BKM:vort} and the
Gevrey regularity was investigated by \citet{KukavicaVicol2009} who
obtained lower bounds on $\delta(t)$ proportional to
$\exp\left(-\int_{0}^t ||\nabla \bds u(\cdot, s)||_{L^\infty}
  \;ds\right)$. The width of the analyticity strip $\delta(t)$ can be
conveniently estimated from the Fourier spectrum of the solution
$\u(t)$ \citep{ssf83} and is often used as a regularity indicator
whose vanishing implies a singularity formation in numerical
computations of Euler flows \citep{bb12}.

\subsection{Statement of the optimization problem} \label{sec:statement:prob}

Motivated by the local existence result in Theorem \ref{thm:Hm},
the goal of our optimization-based formulation is to
construct initial data $\bbeta$ with a prescribed $H^m$, $m > 5/2$,
norm, such that at a given time $T>0$, the corresponding solution
$\u(T; \bbeta)$ of the Euler system \eqref{eq:Euler} has the largest
possible $H^m$ norm. To fix attention, we will hereafter use $m=3$.
Furthermore, since we consider zero-mean initial conditions,
cf.~\eqref{eq:space:initial:condition}, without loss of generality, we
can use the $H^3$ seminorm instead of the full $H^3$ norm as this will make
the optimization problem more similar to the problem considered by
\citet{KangYunProtas2020}.

We define the objective functional $\Phi_T: V \to \mbb R^+$ as
\begin{equation}\label{eq:obj}
\Phi_{T}(\bbeta) := \| \u(T; \bbeta) \|^2_{\dot{H}^3},
\end{equation}
where $\u(T;\bbeta)$ is the solution of the Euler system
\eqref{eq:Euler} obtained with the initial condition $\bbeta \in V$.
System \eqref{eq:Euler} possesses a scaling property such that if
$\left\{ \bds u(\bds x, t), p(\bds x, t) \right\}$ is a solution, then
for any $\lambda > 0$,
\begin{equation}\label{eq:scale}
\left\{ \bds u^\lambda(\bds x, t) := \lambda \bds u(\bds x, \lambda t), \qquad
p^\lambda(\bds x, t) := \lambda^2 p(\bds x, \lambda t) \right\}
\end{equation}
is also a solution. Hence, for any nonzero initial condition {$\bbeta$},  we have the identity
\begin{equation}
\Phi_{T}(\bbeta) =  ||\bbeta||^2_{\dot H^3}\Phi_{||\bbeta||_{\dot H^3}T}\left(\frac{\bbeta}{||\bbeta||_{\dot H^3}}\right).
\end{equation}
Therefore, we can restrict our discussion to initial conditions with
unit $\dot H^3$ seminorm which belong to a closed manifold
$\M_1 \subset V$ defined as
\begin{equation}\label{eq:M1}
\M_1 = \{\bds v \in V:\,  ||\bds v||_{\dot H^3} = 1\}.
\end{equation}
Thus, we  arrive at the following optimization problem
\begin{problem}\label{pb:H3}
Given $T \in \mbb R_+$, find 
\begin{equation}
\teta_{T} = \argmax\limits_{\bbeta \in \M_{1}} \Phi_T(\bbeta).
\end{equation}
\end{problem}
Flows corresponding to optimal initial conditions $\teta_T$ are
referred to as ``extreme''.  Since a priori we do not know if a
singularity may appear in Euler flows on a given time interval
$[0,T]$, we solve the optimization problem for increasing values of
$T$ and deduce whether a singularity may occur inside the interval
$[0,T]$ by performing resolution refinement as discussed in detail in
Section \ref{sec:results}.  If the time window $[0,T]$ falls within the
interval of local existence guaranteed by Theorem \ref{thm:Hm}, then
for any $\bbeta \in \M_1$, we expect $\Phi_T(\bbeta)$ to be finite
such that it will remain bounded upon resolution refinement.  On the
other hand, if there exists an $\bbeta\in \M_1$, which will lead to a
finite-time blow-up inside a sufficiently long time interval $[0, T]$,
we anticipate $\Phi_T(\bbeta)$ to diverge as the resolution is
refined.

\section{Solution approach}
\label{sec:approach}

We solve Problem \ref{pb:H3} using a Riemannian conjugate gradient
approach which is a modification of a gradient ascent method given by
the iterative relation
\begin{equation}\label{eq:GM}
\bbeta_T^{(n+1)} = \bbeta_T^{(n)} + \tau_n \nabla \Phi_T\left(\bbeta_T^{(n)}\right),
\qquad \bbeta_{T}^{(0)} = \bbeta_0, \qquad n=0,1,\dots,
\end{equation}
where $\nabla \Phi_T\left(\bbeta^{(n)}\right)$ is the gradient of the
objective functional $\Phi_T(\bbeta)$ evaluated at the
element $\bbeta_T^{(n)}$, $\tau_n$ is the step size along
the gradient direction and $\bbeta_0$ is the initial guess. A local
maximizer $\teta_T$ of Problem \ref{pb:H3} can then be found as the
limit $\teta_T = \lim_{n\to\infty} \bbeta_T^{(n)}$, such that
\begin{equation}\label{eq:Phi:GM:limit}
\Phi_T\left(\teta_T\right) = \lim_{n\to\infty} \Phi_T\left(\bbeta_T^{(n)}\right)  = \max_{\bbeta \in \M_1}\Phi_{T}(\bbeta) =: \tPhi_{T; \; \bbeta_0}.
\end{equation}
Since our optimization problem is non-convex, the sequence constructed
in \eqref{eq:GM} may converge to different local maximizers depending
on the initial guess $\bbeta_0$. Therefore, we use the subscript
$\bbeta_0$ on the RHS in \eqref{eq:Phi:GM:limit} to
indicate the dependence of the local maximizer on the initial guess.

In the spirit of the {``optimize-then-discretize''} paradigm \citep{g03},
we first formulate our approach in the infinite-dimensional
(continuous) setting and then discretize the resulting relations in
numerical computations.  A key element of the iterative procedure
\eqref{eq:GM} is the evaluation of the gradient $\nabla
\Phi_T\left(\bbeta \right)$, which is based on solving a
suitably-defined adjoint system backwards in time as discussed in
Section \ref{sec:adjoint}.  The Riemannian conjugate gradient method,
which accelerates the gradient ascent method \eqref{eq:GM} for
problems defined on smooth manifolds such as $\M_1$, is introduced in
Section \ref{sec:RCG}, whereas numerical techniques used to discretize the
problem are briefly discussed in Section \ref{sec:num:discretization}.
Finally, different initial guesses we use are listed in
Section \ref{sec:IG} and in Section \ref{sec:diagno} we describe the
diagnostic quantities that will be analyzed.

\subsection{Evaluation of the gradient}
\label{sec:adjoint}

To compute the gradient $\nabla \Phi_T(\bbeta)$ of the objective
functional in \eqref{eq:GM}, we first introduce the G\^{a}teaux
(directional) differential $\Phi_T'(\bbeta; \bbeta'): V \times V
\to \mbb R$
\begin{equation}\label{eq:differential}
\Phi_T'(\bbeta; \bbeta') := \lim_{\epsilon \to 0} \frac{1}{\epsilon}
\left[ \Phi_T\left(\bbeta + \epsilon \bbeta'\right) - \Phi_T(\bbeta)\right],
\end{equation}
which represents the variation of the objective functional
$\Phi_T(\bbeta)$ resulting from applying an infinitesimal perturbation
proportional to $\bbeta' \in V$ to the initial condition $\bbeta$.
Fixing the first argument of $\Phi_T'(\bbeta; \bbeta')$, we can view
the G\^{a}teaux differential \eqref{eq:differential} as a bounded
linear functional on $V$.  Therefore, we can define $\nabla
\Phi_T(\bbeta)$ using the Riesz representation theorem \citep{l69} in
terms of the inner product on $G^\sigma$, cf.~\eqref{eq:G}, as
\begin{equation}\label{eq:dif:grad}
\Phi_T'(\bbeta; \bbeta') = \left\la \nabla \Phi_T(\bbeta), \bbeta'\right\ra_{G^{\sigma}},
\qquad \bbeta, \bbeta' \in V.
\end{equation}
The G\^{a}teaux differential can be evaluated by substituting
\eqref{eq:obj} into \eqref{eq:differential}, which yields
\begin{equation}\label{eq:dif:L2}
\Phi_T'(\bbeta; \bbeta') = 2\left\la\bds u(\cdot, T), \bds u'(\cdot, T)\right\ra_{\dot H^3}
= 2\left\la|D|^6 \bds u(\cdot, T), \bds u'(\cdot, T)\right\ra_{L^2}.
\end{equation}
Here $\bds u'(\bds x, t)$ is the solution of the linearization of the
Euler system \eqref{eq:Euler} around its solution $\bds u(\bds x, t;
\bbeta)$, which is defined by
\begin{equation}\label{eq:L}
\begin{aligned}
\mc L
\begin{bmatrix} \bds{u}' \\ p' \end{bmatrix}
:&=
\begin{bmatrix}
\partial_t \bds u' + \bds u' \cdot \bnabla \bds u + \bds u\cdot \bnabla \bds u' + \bnabla p '\\[3pt]
\bnabla \cdot \bds u'
\end{bmatrix}=
\begin{bmatrix}
\bds 0\\[3pt]
0
\end{bmatrix},\\[3pt]
{\u'(\bds x, 0)} &{= \bbeta'(\bds x)},
\end{aligned}
\end{equation}
where $\bbeta'$ is the perturbation of the initial condition and
$p'(\bds x, t)$ is the corresponding pressure perturbation.

We note, however, that the expression \eqref{eq:dif:L2} for the
G\^{a}teaux differential is not yet consistent with the Riesz
representation form \eqref{eq:dif:grad}, since the perturbation
$\bbeta'$ of the initial condition does not appear in it explicitly,
but is instead ``hidden'' in the initial condition of the linearized
problem \eqref{eq:L}. In order to transform expression
\eqref{eq:dif:L2} into the required Riesz representation form, where
the perturbation $\bbeta'$ appears explicitly as a linear factor, we
introduce the adjoint states $\left\{\bds u^*: \mbb T^3 \times [0, T]
  \to \mbb R^3, \ p^*: \mbb T^3 \times [0, T] \to \mbb R\right\}$ and
the following duality-pairing relation
\begin{equation}\label{eq:pairing}
\begin{aligned}
\left(\mc L 
\begin{bmatrix} \bds{u}' \\ p' \end{bmatrix}, \, 
\begin{bmatrix} \bds{u}^* \\ p^* \end{bmatrix}
\right) : =& 
\int_0^{T}\int_{\mbb T^3} 
\mc L \begin{bmatrix} \bds{u}' \\ p' \end{bmatrix} \cdot 
\begin{bmatrix} \bds{u}^* \\ p^* \end{bmatrix}
\, d\bds x \,dt\\[3pt]
= &\left(
\begin{bmatrix} \bds{u}' \\ p' \end{bmatrix}, \, 
\mc L^* \begin{bmatrix} \bds{u}^* \\ p^* \end{bmatrix}
\right)
+ \int_{\mbb T^3} \bds u'(\bds x, T) \cdot \bds u^*(\bds x, T) \, d\bds x\\[3pt]
&- \int_{\mbb T^3} \bbeta'(\bds x) \cdot \bds u^*(\bds x, 0) \, d\bds x\\[3pt]
= & 0.
\end{aligned}
\end{equation}
Here ``$\cdot$'' denotes the usual Euclidean inner product on
$\RR^4$ and $\mc L^*$ is the {\em adjoint} operator defined
in terms of the following system, which is obtained by performing
integration by parts with respect to both space and time in
(\ref{eq:pairing})
\begin{equation}\label{eq:adjoint}
\begin{aligned}
\mc L^* \begin{bmatrix} \bds{u}^* \\ p^* \end{bmatrix} :&= 
\begin{bmatrix}
-\partial_t \bds u^* - (\bnabla \bds u^* + (\bnabla \bds u ^*)^T)\bds u - \bnabla p^*\\[3pt]
-\bnabla \cdot \bds u^*
\end{bmatrix}
= \begin{bmatrix}
\bds 0\\[3pt]
0
\end{bmatrix}, \\[3pt]
{\bds u^*(\bds x, T)} &{=  2|D|^6\bds u(\bds x, T)}.
\end{aligned} 
\end{equation}
We remark that system \eqref{eq:adjoint} is a {\em terminal-value}
problem and as such has to be integrated backwards in time.  We also
note that the spatial mean of its solutions is conserved
during the time evolution and therefore we have
\begin{equation}
\int_{\mbb T^3} \bds u^*(\bds x, T) \; d\bds x 
=  \int_{\mbb T^3} \bds u^*(\bds x, 0) \; d\bds x = 0.
\end{equation}
 Combining \eqref{eq:dif:grad},
\eqref{eq:dif:L2}, \eqref{eq:pairing} and \eqref{eq:adjoint}, we have
\begin{equation}
\Phi'_T(\bbeta; \bds u'_0)
= \int_{\mbb T^3} \bds u'(\bds x, T) \cdot \bds u^*(\bds x, T) \, d\bds x
= \int_{\mbb T^3} \bds u'_0(\bds x) \cdot \bds u^*(\bds x,0)\, d\bds x
= \left\la \nabla \Phi_T(\bbeta), \bbeta'\right\ra_{G^{\sigma}}.
\end{equation}
Using the definition of the Gevrey inner product \eqref{eq:Gevrey:def}
in the last equality above, we obtain
{
\begin{equation}
\begin{aligned}
\left\la \nabla \Phi_T(\bbeta), \bbeta'\right\ra_{G^{\sigma}} 
& = \int_{\mbb T^3}  \bds u^*(\bds x,0) \cdot \bbeta'(\bds x) \, d\bds x \\
&= \int_{\mbb T^3} (1+|D|^2)^{3}e^{2\sigma|D|} \left[(1+|D|^2)^{-3}e^{-2\sigma|D|} \bds u^*(\bds x, 0)\right] \cdot \bbeta'(\bds x) \; d\bds x \\[3pt]
&= \left\la (1+|D|^2)^{-3}e^{-2\sigma|D|} \bds u^*(\cdot, 0), \bbeta'(\cdot)\right\ra_{G^\sigma},
\end{aligned}
\end{equation}
}
which allows us to identify the gradient as 
\begin{equation}
\nabla \Phi_T(\bbeta) = (1+|D|^2)^{-3}e^{-2\sigma|D|} \bds u^*(0).
\end{equation}
In summary, evaluation of this expression requires the solution of the
adjoint system \eqref{eq:adjoint}, which depends on the solution of
the Euler system \eqref{eq:Euler} with the initial condition $\bbeta$.

\subsection{Riemannian conjugate gradient method}\label{sec:RCG}

For any $\bbeta \in \M_1$, $\bbeta + \tau \nabla \Phi_T(\bbeta)$ given
by \eqref{eq:GM} is divergence-free but does not necessarily have a
unit $\dot H^3$ seminorm and thus will not in general belong to
$\M_1$.  In order to enforce this constraint and accelerate
iterations, we use a Riemannian conjugate gradient method that
consists of three steps: first, we project the gradient $\nabla
\Phi_T(\bbeta)$ onto the tangent space to $\M_1$ at
$\bbeta$, then perform a suitable {\em vector transport} operation in
order to construct a Riemannian conjugate ascent direction using the
previous search direction and, finally, retract the resulting state
back to the constraint manifold $\M_1$. An analogous approach
  was employed by \citep{DanailaProtas2017} to solve an
  infinite-dimensional minimization problem with a similar structure
  arising in quantum fluids.

\begin{figure}
\centering
     \includegraphics[width=0.6\textwidth]{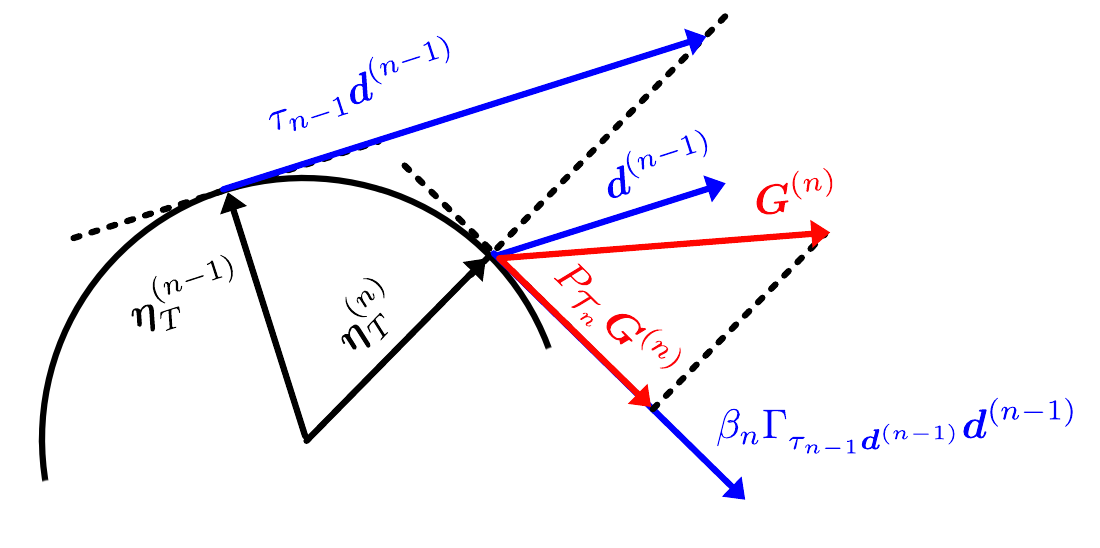}
     \caption{Schematic illustration of the Riemannian
         conjugate gradient method \eqref{eq:RCG}.}
        \label{fig:RCG}
     \end{figure}

We first define the tangent space to $\M_1$ at $\vv$ as
\begin{equation}\label{eq:tangent:sigma}
\mc T_{\bds v} \M_1 = \Big\{
\bds z \in V:\,
\la \bds v, \bds z\ra_{\dot H^3}= 0
\Big\},
\end{equation}
where the last condition represents the G\^{a}teaux differential of
the constraint $\| \vv \|_{\dot{H}^3} = 1$. For any $z \in \mc T_{\bds
  v} \M_1$, this condition can be expressed as
\begin{equation}
\begin{aligned}
\la \bds v, \bds z\ra_{\dot H^3} &= \int_{\mbb T^3} |D|^3 \bds v \cdot |D|^3 \bds z \, d\bds x\\[3pt]
&= \int_{\mbb T^3} (1+|D|^2)^3e^{2\sigma|D|} \Big((1+|D|^2)^{-3}e^{-2\sigma|D|}|D|^6\bds v\Big)\cdot \bds z \, d\bds x\\[3pt]
 &= \left\la (1+|D|^2)^{-3}e^{-2\sigma|D|}|D|^6\bds v,\bds z \right\ra_{G^{\sigma}} = 0.
 \end{aligned}
\end{equation}
Thus, at each point $\vv \in \M_1$, the tangent space $\mc T_{\bds v}
\M_1$ can be characterized using the unit ``normal'' vector given by
\begin{equation}
\bds n_{\bds v} = \frac{(1+|D|^2)^{-3}e^{-2\sigma|D|}|D|^6\bds v}{\big|\big|(1+|D|^2)^{-3}e^{-2\sigma|D|}|D|^6\bds v\big|\big|_{G^{\sigma}}}
\label{eq:n}
\end{equation}
which allows us to construct the projection operator $\PP_{\mc T_{\vv}
  \M_1} \; : \; V \rightarrow {\mc T_{\vv} \M_1}$, cf.~figure
\ref{fig:RCG},
\begin{equation}\label{eq:proj:tangent}
\forall \w \in V, \qquad \PP_{\mc T_{\bds v} \M_{1}} \bds w = \bds w - \left\la \n_{\vv}, \bds w \right\ra_{G^{\sigma}}  \n_{\vv}.
\end{equation}
In addition, we also need to introduce the retraction operator $\PR \;
: \; V \rightarrow \M_1$ which has the form
  of normalization needed to satisfy the constraint $||\bds v||_{\dot
    H^3} = 1$ \citep{ams08,Sato:2021:RCG}
\begin{equation}
\PR(\bds v) = \frac{\bds v}{||\bds v||_{\dot H^3}}, \qquad \bds v\neq \bds 0.
\label{eq:R} 
\end{equation}

In the standard (Euclidean) conjugate gradient approach, the search
direction at $(n+1)$-th iteration is obtained as a suitable linear
combination of the steepest ascent direction (i.e., the gradient)
computed at $(n+1)$-th iteration and the search direction obtained at
$n$-th iteration. In the Riemannian setting the difficulty with this
approach is that these two directions do not belong to the same linear
space and hence cannot be added directly. In order to circumvent this
difficulty, we introduce the {\em vector transport} operation defined
as
\begin{equation}
{{\cal TM}_1 \oplus {\cal TM}_1 \rightarrow {\cal TM}_1 \; : \; 
(\bds \varphi,\bxi) \longmapsto \Gamma_{\bds \varphi}(\bxi) \in {\cal TM}_1,}
\label{eq:T}
\end{equation}
where ${\cal TM}_1 = \cup_{\bds v \in {\cal M}_1} {{\cal T}_{\bds v}
  {\cal M}_1}$ is the tangent bundle on $\M_1$, describing how the
vector field $\bxi$ is transported along the manifold $\M_1$ in the
direction of $\bds \varphi$ \citep{ams08}.  The vector transport
provides a map between the tangent spaces ${{\cal T}_{\bbeta_T^{(n)}}
  {\cal M}_1}$ and ${{\cal T}_{\bbeta_T^{(n+1)}} {\cal M}_1}$ obtained
at two consecutive iterations, so that algebraic operations can be
performed on vectors belonging to these subspaces.  It therefore
generalizes the concept of the parallel translation to the motion on
the manifold.  In general, vector transport is defined up to a
multiplicative prefactor and can be computed via the differentiated
retraction as \citep{ams08}
\begin{equation}
  \T_{\bds\varphi_{\u}}(\bxi_{\u}) = \frac{d}{dt} \PR_{\u}({\bds\varphi_{\u} + t \bxi_{\u}}) \big|_{t=0} = 
  \frac{1}{\| \u + \bds\varphi_{\u} \|_{\dot H^3}}\left[ \bxi_{\u} - \frac{\langle \u+\bds\varphi_{\u}, \bxi_{\u} \rangle_{\dot H^3}}{\| \u + \bds\varphi_{\u} \|_{\dot H^3}^2} (\u+\bds\varphi_{\u}) \right], \quad \u \in \M_1
  \label{eq:T1}
\end{equation}

Combining the elements introduced above, the Riemannian conjugate
     gradient approach is given by the iterative relation
\begin{equation} \label{eq:RCG}
\bbeta_T^{(n+1)}  = \PR\Big[\bbeta_T^{(n)} + \tau_n \bds d^{(n)}\Big], \qquad \bbeta_T^{(0)} = \bbeta_0, \qquad n=0,1,\dots,
\end{equation}
and is also schematically illustrated in figure
  \ref{fig:RCG}.  Here the search direction $\bds d^{(n)}$ is
computed as
\begin{equation}
\begin{aligned}
\bds d^{(0)} &= \PP_{\mc T_0} \bds G^{(0)},\\
\bds d^{(n)} &= \PP_{\mc T_n} \bds G^{(n)} + \beta_n \Gamma_{\tau_{n-1} \d^{n-1}}\left(\d^{n-1}\right),
\qquad n\geq 1,
\end{aligned}
\end{equation}
where $G^{(n)} := \nabla\Phi_T\left(\bbeta^{(n)}\right)$ and $\mc T_n
:= \mc T_{\bbeta^{(n)}} \M_1$.  The ``momentum'' term $\beta_n$ is
chosen to enforce the conjugacy of consecutive search directions and
is computed using the Polak-Ribi\`ere approach \citep{nw00}
\begin{equation}
\beta_n = \frac{\Big\la \PP_{\mc T_n} \bds G^{(n)}, 
\Big( \PP_{\mc T_n} \bds G^{(n)} -
\Gamma_{\tau_{n-1} \d^{n-1}}\PP_{\mc T_{n-1}} \bds G^{(n-1)}\Big)\Big\ra_{G^{\sigma}}}
{\Big|\Big|  \PP_{\mc T_{n-1}} \bds G^{(n-1)}\Big|\Big|^2_{G^{\sigma}}}.
\end{equation}
The step size $\tau_n$ in \eqref{eq:RCG} is determined by solving
the arc-search problem
\begin{equation}
\tau_n = \argmax_{\tau > 0} \left\{\Phi_{T} 
\left(\PR\Big[\bbeta^{(n)} + \tau\bds d^{(n)}\Big]\right)  \right\}
\label{eq:arcsearch}
\end{equation}
using a suitable derivative-free approach, such as a variant of
 Brent's algorithm ~\citep{numRecipes}. Due to the presence of the
retraction operator \eqref{eq:R} in \eqref{eq:RCG}, the search in
\eqref{eq:arcsearch} is performed following a geodesic arc on the
manifold $\M_1$, this problem can be regarded as a generalization of
the more common line search approach \citep{numRecipes}.

\subsection{Numerical methods} \label{sec:num:methods}

\subsubsection{Discretization, validation and numerical parameters} \label{sec:num:discretization}

In numerical computations, systems \eqref{eq:Euler} and
\eqref{eq:adjoint} are discretized in space using a standard
pseudospectral Fourier method where derivatives are evaluated in the
Fourier space, whereas nonlinear products are computed in the physical
space on an uniform grid with $N$ points in each direction. As regards
dealiasing, we {use the Gaussian filter proposed by \citet{hl07}
  which is defined in terms of the Fourier multiplier $\widehat
  G_{\bds j} := \exp \left\{-36[(2|j_1|/N)^{36} + (2|j_2|/N)^{36} +
    (2|j_3|/N)^{36}]\right\}$. As discussed in Section
  \ref{sec:Tlong}, this filter will also serve to prevent the blow-up
  of the numerical solution after it becomes under-resolved.}  The
resulting system of ordinary differential equations is discretized in
time using an explicit fourth-order Runge-Kutta (RK4) method. The code
is parallelized using MPI and the fast Fourier transforms (FFTs) are
computed using the parallel version of the software library FFTW
\citep{FFTW}. We consider four spatial resolutions with $N^3 = 128^3,
256^3, 512^3, 1024^3$ and the time steps $\Delta t$ are adjusted to
satisfy the stability conditions of the RK4 method.  If needed, the
resolution used will be denoted with a superscript $N$, i.e.,
$\u^N(t)$ represents the numerical solution of the Euler system
\eqref{eq:Euler} computed with the resolution $N^3$, etc. At each
iteration in \eqref{eq:RCG}, the solution $\u(t)$ of the Euler system
\eqref{eq:Euler} is needed to construct coefficients in the adjoint
system \eqref{eq:adjoint}.  For the three smaller resolutions, we
store the solutions of \eqref{eq:Euler} at discrete time steps so that
they can be read directly from files when solving \eqref{eq:adjoint}.
However, this is not possible for the resolution of $1024^3$ due to
storage limitations. We get around this difficulty using the
time-reversibility of the Euler system \eqref{eq:Euler} to evolve this
system backwards in time while solving the adjoint system; this allows
us to determine the coefficients in the latter system without having
to store the entire solution to the former.

The solver for the Euler system \eqref{eq:Euler} was validated by
verifying the convergence of the numerical solutions to certain exact
solutions of \eqref{eq:Euler} as we increase the spatial resolution
and/or decrease the time step.  As regards the accuracy of
the evaluation of the gradient $\nabla\Phi_T\left(\bbeta \right)$, we
consider the quantity \citep{a14}
\begin{equation}
\kappa^N(\epsilon) := \frac{\Phi_T^N\left(\bbeta + \epsilon\bbeta'\right) - \Phi_T^N(\bbeta)}
{\epsilon \left\la \nabla \Phi_T^N(\bbeta), \bbeta'\right\ra_{G^\sigma}}
\label{eq:kappa}
\end{equation}
which is the ratio of a finite-difference approximation of the
G\^{a}teaux differential \eqref{eq:differential} and its
representation in terms of the Riesz representation formula \eqref{eq:dif:grad}, both
approximated using the resolution $N^3$. In theory, we expect that
$\lim_{N \rightarrow \infty}\lim_{\epsilon \rightarrow 0}
\kappa^N(\epsilon) = 1$. However, in numerical computations performed
with finite-precision arithmetics, $\kappa^N(\epsilon)$ will become
unbounded as $\epsilon \rightarrow 0$ due to subtractive cancellation
errors in the numerator in \eqref{eq:kappa}. For intermediate values
of $\epsilon$, the quantity $|\kappa^N(\epsilon) - 1|$ is a measure of
inaccuracies involved in the computation of $\Phi_T^N\left(\bbeta
\right)$ and $\nabla\Phi_T^N\left(\bbeta \right)$ due to approximation
errors in the numerical solution of systems \eqref{eq:Euler} and
\eqref{eq:adjoint}. In our validation tests, we observed that, as
expected, $|\kappa^N(\epsilon) - 1|$ is reduced as the discretization
parameters $N$ and $\Delta t$ are refined.

In computations with the nonlinear conjugate gradient method, it is
common to restart the algorithm periodically with a simple gradient
step, i.e., by setting $\beta_n = 0$, which allows us to improve
convergence by getting rid of ineffectual old information
\citep{nw00}. In our computations we restart the algorithm
\eqref{eq:RCG} based on the following two criteria: \medskip
\begin{itemize}
\item[(1)] \, $n = 20 k$, $k \in \mbb Z_+$,
\smallskip
\item[(2)] \, The search direction $\bds d^{(n)}$ fails to be an ascent direction, i.e., 
\begin{equation}
\frac{\Big\la \bds d^{(n)}, \PP_{\mc T_n} G^{(n)}\Big\ra_{G^{\sigma}}}
{\Big|\Big| \bds d^{(n)}\Big|\Big|_{G^{\sigma}}\Big|\Big| \PP_{\mc T_n} G^{(n)}\Big|\Big|_{G^{\sigma}}}
< \emach, \qquad 0 < \emach \ll 1.
\end{equation}

\end{itemize}
Iterations \eqref{eq:RCG} are declared converged when the relative
change of the objective functional between two consecutive iterations
becomes smaller than a specified tolerance $0 < \tol \ll 1$, i.e.,
when
\begin{equation}
{0 \leq \frac{\tPhi_T^N\left(\bbeta_T^{(n+1)}\right) - \tPhi_T^N\left(\bbeta_T^{(n)}\right)}{\tPhi_T^N\left(\bbeta_T^{(n)}\right)}
< \tol.}
\end{equation}

\medskip

\subsubsection{Initial guesses} \label{sec:IG}

Since Problem \ref{pb:H3} is non-convex, iterations \eqref{eq:RCG} may
produce local maximizers which will depend on the initial guess
$\bbeta_0$.  In numerical computations, we thus consider a variety of
different initial guesses listed below, all normalized such that $\|
\bbeta_0\|_{\dot{H}^3} = 1$. For simplicity, we use the same notation
for the original initial conditions and their normalizations.  In each
case, we solve Problem~\ref{pb:H3} to obtain an optimal initial
condition $\teta_T$ such that for any given $T$ the norm $\|
\u(T;\teta_T) \|_{\dot{H}^3}$ is larger than $\| \u(T;\bbeta_0)
\|_{\dot{H}^3}$ where $\bbeta_0$ is each of the following
initial guesses.

\smallskip
\begin{itemize}
\item[(1)] \  The 3D Taylor-Green vortex \citep{tg37}:
\begin{equation}\label{eq:TG}
\bbeta_{\TG} := 
\begin{bmatrix}
\sin(2\pi x) \cos(2\pi y) \cos(2\pi z)\\
-\cos(2\pi x)\sin(2\pi y) \cos(2\pi z)\\
0
\end{bmatrix}.
\end{equation}
The Taylor-Green vortex has been widely used as a candidate for
potential blow-up in Euler flows, however, it is still an open
question whether this initial condition can indeed lead to a
singularity formation in a finite time \citep{cb05,bb12,Fehn2022}.  We
mention that the time windows $[0,T]$ considered in this paper are
much shorter than the times when a potentially singular behavior was
observed in these earlier studies.

\smallskip
\item[(2)] \  Random initial condition:
\begin{equation}
{\bbeta_{\rand}} := \PP_{\text{L}} {\bds v_\rand} := {\bds v_\rand} - \bnabla \Delta^{-1}(\bnabla\cdot {\bds v_\rand}),
\label{eq:random}
\end{equation}
where $\PP_{\text{L}}$ is the Leray projector \citep{mb02} with the
property that for any $\w \in H^1(\TT^3)$, $\bnabla\cdot\left(
  \PP_{\text{L}}\w \right) = 0$, and $\bds v_\rand$ is given by
\begin{equation}
{\bds v_\rand} = \begin{bmatrix}
v_1 \\
v_2 \\
v_3
\end{bmatrix}
= \begin{bmatrix}
\sum_{|j_1| + |j_2| + |j_3| \leq N_0} e^{-|\bds j|/(4\pi)}e^{i2\pi\theta_1(j_1)}e^{i2\pi \bds j\cdot \bds x} + c.c.\\[3pt]
\sum_{|j_1| + |j_2| + |j_3| \leq N_0} e^{-|\bds j|/(4\pi)}e^{i2\pi\theta_2(j_2)}e^{i2\pi \bds j\cdot \bds x} + c.c.\\[3pt]
\sum_{|j_1| + |j_2| + |j_3| \leq N_0} e^{-|\bds j|/(4\pi)}e^{i2\pi\theta_3(j_3)}e^{i2\pi \bds j\cdot \bds x} + c.c.
\end{bmatrix},
\end{equation}
in which $\theta_1$, $\theta_2$ and $\theta_3$ are $N_0$-dimensional
random variables with uniform distributions on $[0, 1]^{N_0}$. In
practice, we choose $N_0 = 64$ and emphasize that in contrast to the
unimodal Taylor-Green vortex \eqref{eq:TG}, the random initial
condition \eqref{eq:random} has energy distributed over Fourier
coefficients with wavevectors within the shell of radius $2\pi N_0$.

\smallskip
\item[(3)] \ Kerr's initial condition \citep{k93,HouLi2006}:
\begin{equation}
{\bbeta_\Kerr} := \bnabla\times(|D|^{-2} \bds \omega_{\Kerr})
\label{eq:Kerr}
\end{equation}
which represents two perturbed anti-parallel vortex tubes located
symmetrically with respect to the $xy$-plane such that
$\bomega_{\Kerr} (x, y, z) = -\bomega_{\Kerr} (x, y, -z)$.  The
vorticity of the vortex tube above the $xy$-plane is given by
\begin{equation}
\bds \omega_{\Kerr} = 8G\omega(r)[\omega_1, \omega_2, \omega_3]^T,
\end{equation}
where $G$ is a Fourier filter defined by $\hat G_{\bds j} =
\exp\left(-0.05\left(j_1^4+j_2^4+j_3^4\right)\right)$ and
\begin{equation}
\begin{gathered}
{\omega(r) =\begin{cases}
 \exp\left[ \frac{-r^2}{1-r^2} + r^4(1+r^2+r^4)\right], & r < 1,\\
 0, & r \geq 1,
 \end {cases}} 
 \quad r = \frac{1}{R}\sqrt{[x-x(s)]^2 + [z-z(s)]^2}, \\[3pt]
x(s) = \delta_x \cos(\pi s/L_x), \quad 
z(s) = z_0, \\[3pt]
s(y) = y_2 + L_y\delta_{y_1} \sin(\pi y_2/L_y), \quad 
y_2(y) = 4\pi y + L_y\delta_{y_2}\sin(\pi (4\pi y)/L_y),\\[3pt]
\omega_1 = -\frac{\pi\delta_x}{L_x} \left[ 1 + \pi \delta_{y_2} \cos(\pi(4\pi y)/L_y)\right]\cdot
\left[ 1 + \pi \delta_{y_1} \cos(\pi y_2/L_y)\right] \sin(\pi s/L_x), \\[3pt]
\omega_2 = 1, \qquad \qquad 
\omega_3 = 0.
\end{gathered}
\end{equation}
We choose the same parameters as \cite{HouLi2006}, i.e.,
\begin{equation}
\delta_{y_1} = 0.5, \;
\delta_{y_2} = 0.4, \;
\delta_x = -1.6, \;
z_0 = 1.57, \;
R = 0.75, \;
L_x = L_y = 4\pi, \;
L_z = 2\pi.
\end{equation}
We would like to point out that in \cite{HouLi2006}, the equations
are posed on $[-2\pi, 2\pi]^3$ rather than $[-1/2, 1/2]^3$.  
We also note that due to a different normalization, the times up to which
\citet{k93,HouLi2006} were able to compute
solutions of the Euler system~\eqref{eq:Euler} with the initial 
condition $\bbeta_\Kerr$ correspond to $t_1 =
108231$ and $t_2 = 110059$, respectively, and
are significantly longer than the optimization
windows considered in this study.

\item[(4)] \ Hou's axisymmetric initial condition \citep{Hou:22:Euler}: 
\begin{equation}
\bbeta_\Hou := G u_{\theta}\bds e_{\theta},
\label{eq:Hou}
\end{equation}
where $\bds e_{\theta}$ is {the} unit vector in the azimuthal direction of the cylindrical coordinate system and $u_{\theta}$ is the angular velocity component given by
\begin{equation}
u_{\theta} = \begin{cases}
r\exp\left(-r^2/(1-r^2)\right)\frac{12000(1-r^2)^{18}\sin(2\pi z)}{1+12.5\sin^2(\pi z)}, & r < 1,\\
0, & r \geq 1,
\end{cases}
\quad r = \frac{\sqrt{x^2 + y^2}}{0.9}.
\end{equation}
This initial condition is a ``compactified'' version of the initial condition originally considered by \citet{Hou:22:Euler}, where it was defined on a cylindrical domain of unit radius and with a no-flow boundary condition. The results reported by \citet{Hou:22:Euler} indicate that the Euler flow on such a bounded domain obtained  with the initial condition \eqref{eq:Hou} develops a singularity at the origin after a finite, albeit quite long, time.


\end{itemize}
\smallskip

\subsubsection{Diagnostic quantities} \label{sec:diagno}

We analyze numerical solutions of system \eqref{eq:Euler} based on the
following diagnostic quantities: the $\dot H^3$ {seminorm} of the
velocity field $\u(t)$, the $L^\infty$ norm of the vorticity field
$\bomega(t)$ and the width $\delta(t)$ of the analyticity strip
characterizing the velocity field, all of which are functions of time
$t$.

We approximate $\|\bds u(t)\|_{\dot H^3}$ and  $\|\bomega(t)\|_{L^\infty}$ as
\begin{subequations}
\begin{align}
\|\bds u(t)\|_{\dot H^3} &\approx \sqrt {\sum_{j_1 = 0}^{N-1}\sum_{j_2 = 0}^{N-1}\sum_{j_3 = 0}^{N-1}
|2\pi\bds j|^6 \left|\hat{\bds u}_{\bds j}(t)\right|^2},  \\[3pt]
\|\bomega(t)\|_{L^\infty} &\approx \max_{0\leq j_1, j_2,j_3 < N} 
\left|\bomega\left(j_1/N, j_2/N, j_3/N, t\right)\right|.
\end{align}
\end{subequations}

The width of the analyticity strip $\delta(t)$ has been used to
diagnose singularity formation in solutions of various PDEs
\citep{ssf83}, including the 3D Euler {equations} \eqref{eq:Euler}
\citep{bb12}. To approximate it, we first define the energy spectrum
of $\bds u(\bds x, t)$ as
\begin{equation}
e(k, t) := \frac{1}{2}\sum_{j \leq |\bds j| < j+1} |\hat{\bds u}_{\bds j}(t)|^2, \qquad  k = 2\pi j, \quad 
j \in \mbb N,
\label{eq:Ek}
\end{equation}
such that the kinetic energy of the flow is
\begin{equation}
E(t) := \frac{1}{2}\int_{\mbb T^3} |\bds u(\bds x, t)|^2 \, d \bds x 
= \frac{1}{2}\sum_{\bds j\in\mbb Z^3} |\hat{\bds u}_{\bds j}(t)|^2
= \sum_{j} e(2\pi j, t).
\end{equation}
When $\u(t)$ is real-analytic and admits an extension to the complex
{space} $\CC^3$, it can be shown that the energy spectrum \eqref{eq:Ek}
has the following asymptotic representation as $k \rightarrow \infty$
\citep{CarrierKrookPearson2005}
\begin{equation}
e(k, t) = C(t) k^{-n(t)}e^{-2k\delta(t)}, 
\label{eq:Ekinf}
\end{equation}
where $C(t)$ is a scaling constant and $n(t)$ is the order of the
singularity nearest to the real line. Clearly, the loss of analyticity
and hence the singularity formation are signalled by the vanishing of
$\delta (t)$.  Fixing the time $t$ and computing the logarithm of
\eqref{eq:Ekinf}, the parameters $C(t)$, $n(t)$ and $\delta(t)$ can be
determined by performing a least-squares fit to minimize the error
functional
\begin{equation}
\chi(t) := \sum_{k = k_<}^{k_>} \left[\ln e(k,t)-\ln C(t) + n(t) \ln k + 2k\delta(t)\right]^2,
\label{eq:chi}
\end{equation}
where {$k_< = 4\pi$ and $k_>$ is determined in a slightly
  different manner for different time windows $T$ (details are
  provided below).  Since $(\ln (k+2\pi) - \ln k)$ decreases as $k$
  grows, the transformed wavenumbers $\{\ln k, k \in 2\pi\mathbb N\}$
  are not evenly distributed over the interval $[\ln k_<, \ln k_>]$.
  This makes it difficult to accurately determine the parameter
  $n(t)$, as there are fewer grid points for smaller values of $k$. In
  order to get around this problem, we perform the fit using grid
  points distributed uniformly over the interval $[\ln k_<, \ln k_>]$
  where the error functional \eqref{eq:chi} is evaluated by performing
  a linear interpolation of $e(k,t)$ onto these equispaced points. }

\section{Numerical results} 
\label{sec:results}

To distinguish between regular and potentially singular evolution, we
solve Problem \ref{pb:H3} on ``short'' and ``long'' time intervals
$[0, T]$ and perform a resolution refinement, where optimal initial
conditions $\teta_T^N$ and the corresponding Euler flows
$\u^N\left(t;\teta_T^N\right)$ are computed with increasing numerical
resolutions $N^3$, cf.~Section \ref{sec:num:discretization}.  {We
  add that, based on the characteristic length scale $L = 1$ (given by
  the size of the domain $\Omega$) and the ``size'' of the initial
  data $U \sim 1$, the characteristic time scale of the flow is $L/U
  \sim 1$.}


When $T$ is
sufficiently small, such that the Euler system~\eqref{eq:Euler} is
guaranteed by Theorem \ref{thm:Hm} to be well-posed on $[0,T]$, we
expect that
\begin{equation}
\lim_{N\to\infty} \Phi_{T}\left(\teta_T^N\right) < \infty,
\label{eq:wellposed} 
\end{equation}
i.e., the objective functional \eqref{eq:obj} must remain finite upon
resolution refinement.

On the other hand, when $T$ is large enough such that Theorem
\ref{thm:Hm} does not assert the existence of classical solutions on
$[0,T]$ for regular initial data $\bbeta$, the divergence of the
objective functional \eqref{eq:obj} upon resolution refinement
\begin{equation}
\lim_{N\to\infty} \Phi_{T}  \left(\teta_T^N\right)= \infty
\label{eq:blowup} 
\end{equation}
would signal the possibility of a singularity formation at some $t \in
[0,T]$. As we shall see in Section \ref{sec:Tshort} and
Section \ref{sec:Tlong} below, behaviors consistent with
\eqref{eq:wellposed} and \eqref{eq:blowup} are in fact observed in
solutions of Problem \ref{pb:H3} obtained with $T = 25$ and $T=75$,
respectively, with the caveat that the limits in
\eqref{eq:wellposed}--\eqref{eq:blowup} are extrapolated from four
resolutions only ($N^3 = 128^3, 256^3, 512^3, 1024^3$).

\subsection{Results for $T = 25$} 
\label{sec:Tshort}

\begin{figure}
  \centering
\mbox{\subfigure[]{\includegraphics[width=0.48\textwidth]{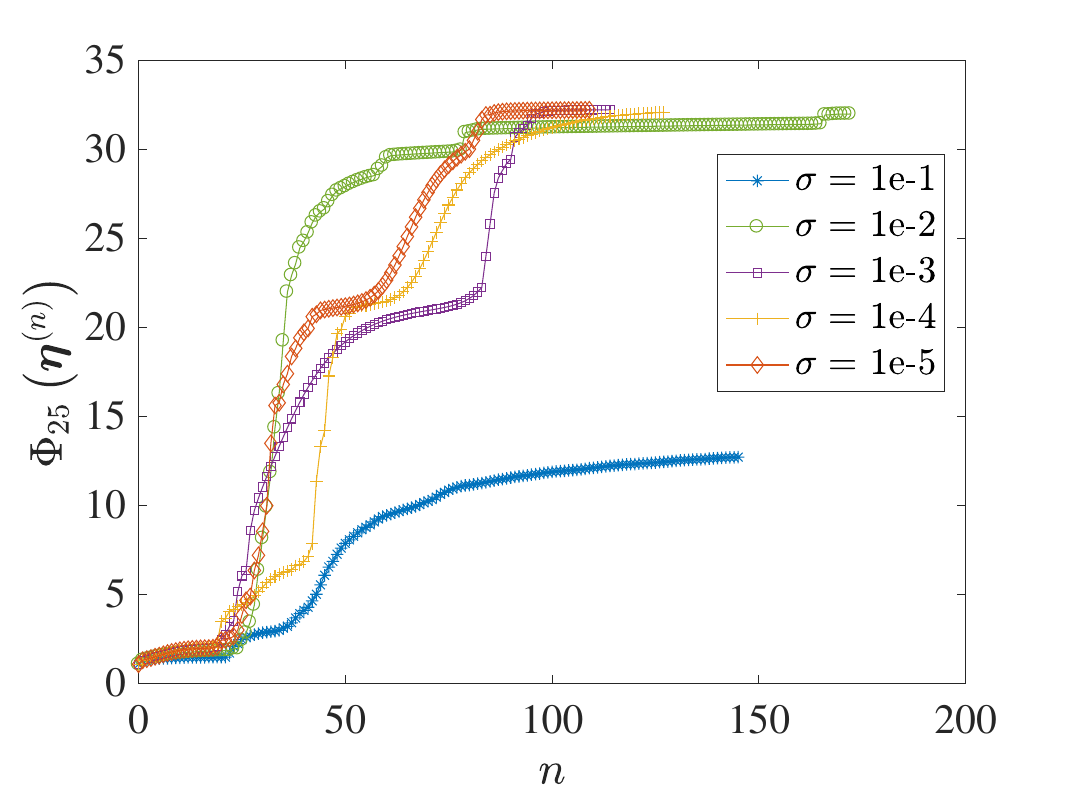}}
\hfill
\subfigure[]{\includegraphics[width=0.48\textwidth]{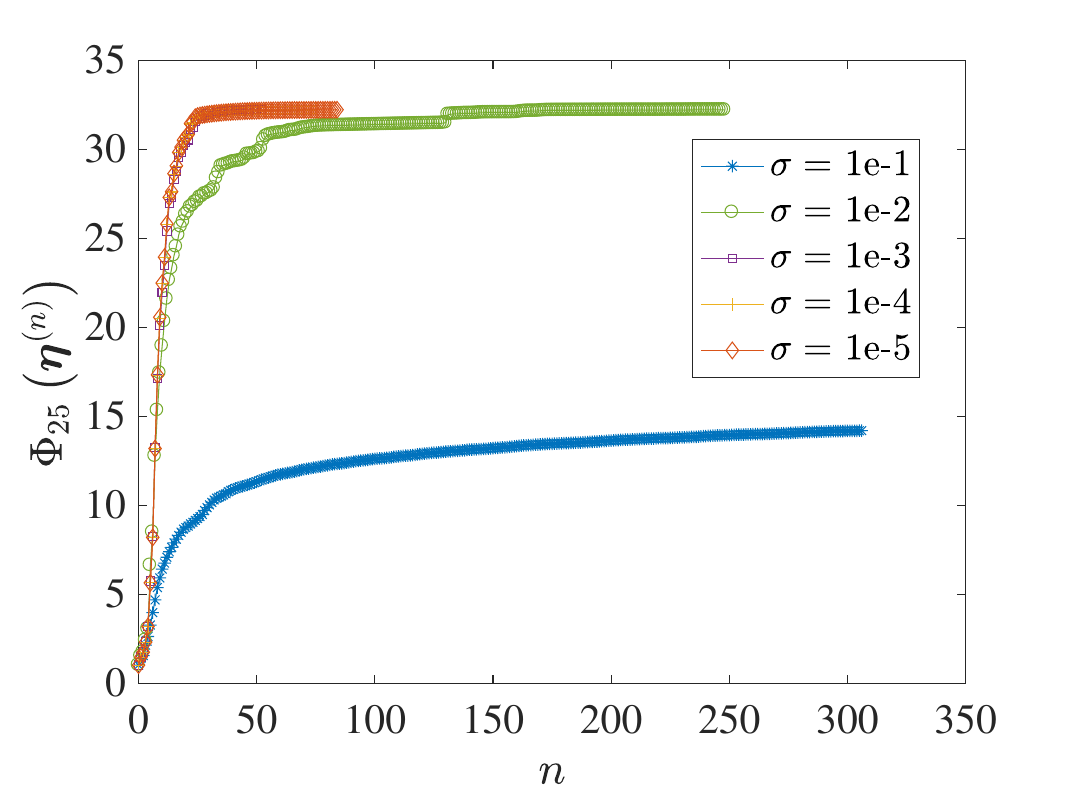}}}
  \caption{[Short time window, $T = 25$] Dependence of the objective
    functional $\Phi_{25}\left(\bbeta^{(n)}\right)$ on the iteration
    index $n$ for different values of $\sigma$ when (a) $\bbeta_{\TG}$
    and (b) $\bbeta_{\rand}$ are used as initial guesses in
    \eqref{eq:RCG}.}
  \label{fig:sigma:compare}
\end{figure}

Since solutions of Problem \ref{pb:H3} depend on the value of the
parameter $\sigma$ defining the Gevrey space via \eqref{eq:Gevrey:def}
and \eqref{eq:G}, we begin by examining the effect of this parameter
on these solutions. We solve Problem \ref{pb:H3} in different Gevrey
spaces $G^\sigma$ with $\sigma = 10^{-1}, 10^{-2}, \ldots, 10^{-5}$
using the resolution {$128^3$} and two initial guesses $\bbeta_\TG$
and $\bbeta_\rand$. The dependence of the value of the objective
functional {$\Phi_{25}\left(\bbeta^{(n)}\right)$} on the iteration
index $n$ in~\eqref{eq:RCG} is shown for different $\sigma$ in figure
2 for the two initial guesses. We observe that as long as $\sigma$ is
sufficiently small ($\sigma < 10^{-1}$), its value does not have an
appreciable effect on the maximum attained value $\tPhi_{25}$ of the
objective functional. However, this parameter does affect the rate of
convergence {and} smaller values of $\sigma$ typically {result} in
faster convergence. This can be interpreted in terms of an ``effective
dimension'' of the discrete space {used to} approximate the optimal
initial condition {$\teta_{25}^N$}, which is larger for smaller
$\sigma$ since decreasing the value of this parameter reduces the
penalty on the Fourier components of {$\teta_{25}^N$} with large
wavenumbers $|\bk|$.  Hereafter we will use $\sigma = 10^{-5}$ which
results in both larger attained values of $\tPhi_{25}$ and faster
convergence of iterations in \eqref{eq:RCG}, cf.~{figure
  \ref{fig:sigma:compare}}.

\begin{figure}
  \centering
\mbox{\subfigure[]{\includegraphics[width=0.48\textwidth]{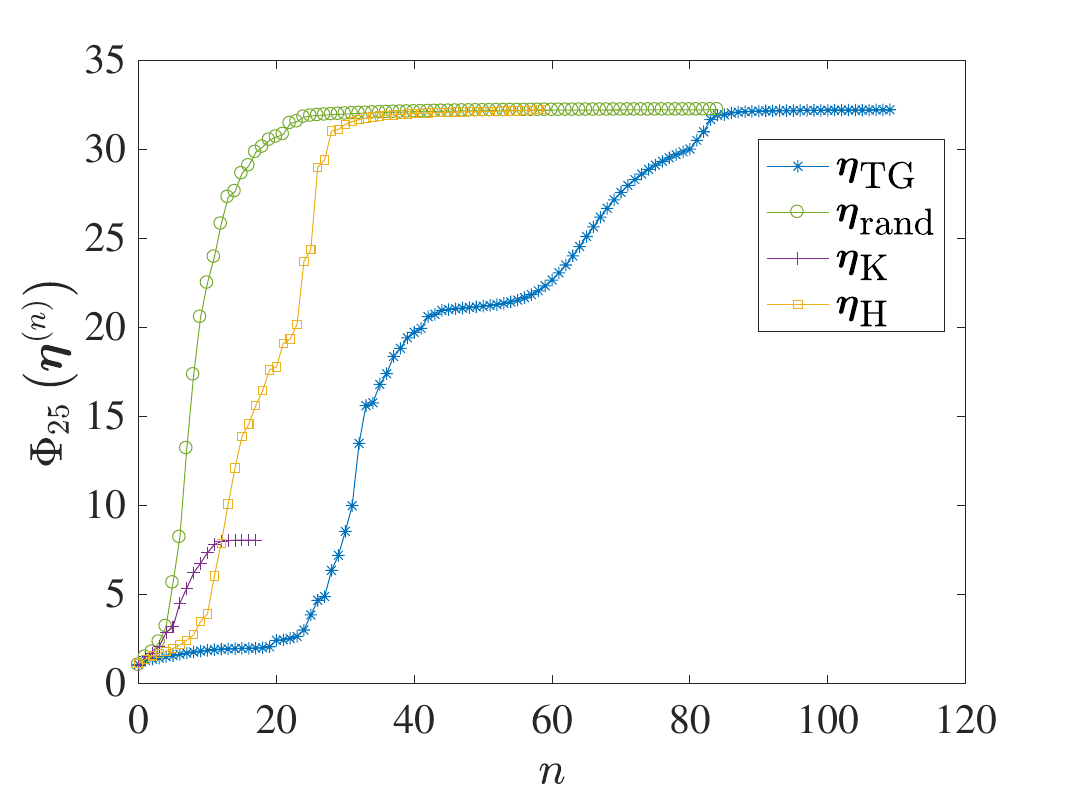}}
\hfill
\subfigure[]{\includegraphics[width=0.48\textwidth]{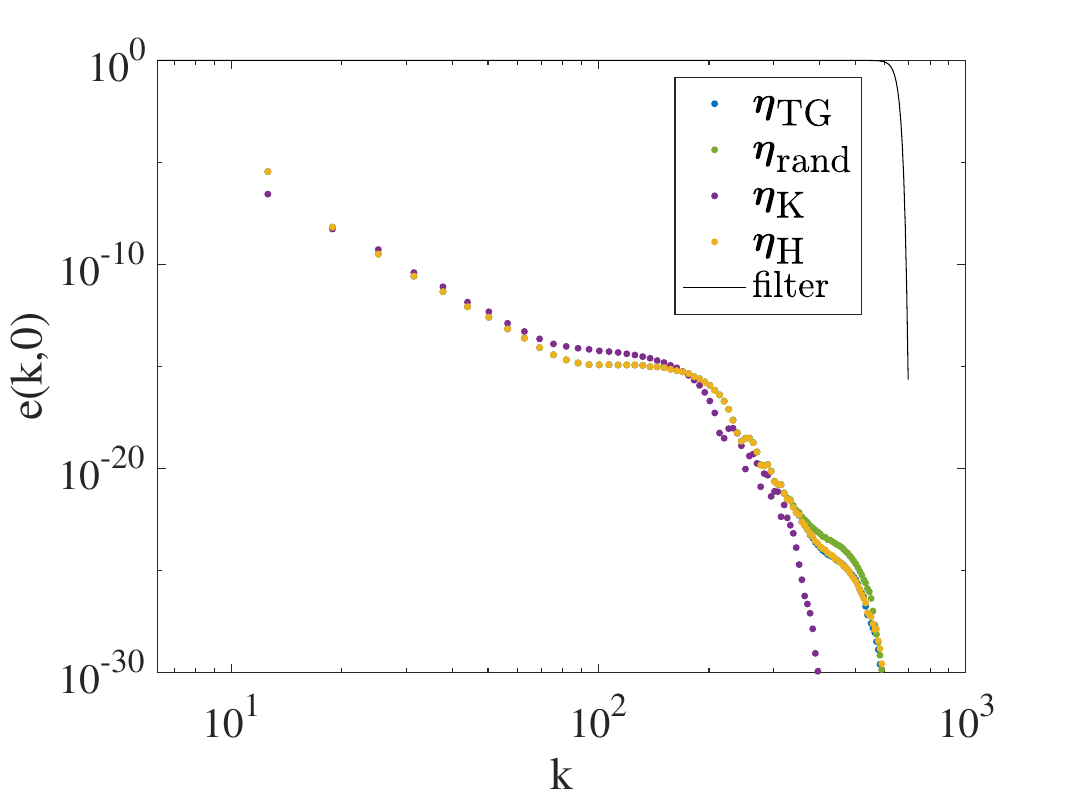}}}
  \caption{[Short time window, $T = 25$] (a) Dependence of the
    objective functional $\Phi_{25}\left(\bbeta^{(n)}\right)$ on the
    iteration index $n$ and (b) energy spectra \eqref{eq:Ek} of the
    optimal initial conditions $\teta_{25}$ corresponding to different
    initial guesses $\bbeta_{\TG}$, $\bbeta_{\rand}$, $\bbeta_{\Kerr}$
    and $\bbeta_{\Hou}$ used in iterations \eqref{eq:RCG}.
    In panel (b) the solid line represents the Gaussian filter we use
    \citep{hl07}.}
  \label{fig:init:25}
\end{figure}

Since Problem \ref{pb:H3} is non-convex, iterations \eqref{eq:RCG} may
lead to different local maximizers depending on the choice of the
initial guess $\bbeta$. To assess this possibility, we solve
Problem~\ref{pb:H3} using {four initial guesses $\bbeta_\TG$,
  $\bbeta_\rand$, $\bbeta_\Kerr$ and $\bbeta_\Hou$} introduced in
Section \ref{sec:IG} {with the resolution $128^3$}. {The results} are
presented in figures \ref{fig:init:25}(a) and \ref{fig:init:25}(b)
where we show, respectively, the value of the objective functional
{$\Phi_{25}\left(\bbeta^{(n)}\right)$} as {a} function of the
iteration index $n$ and the energy spectra \eqref{eq:Ek} of the
optimal initial conditions {$\teta_{25}$} obtained {with different
  initial guesses}.  From the coincidence of the terminal values
$\tPhi_{25}$ in figure~\ref{fig:init:25}(a) and of the energy spectra
$e(k,0)$ in figure \ref{fig:init:25}(b) we conclude that an
essentially the same optimal initial condition {$\teta_{25}$} is found
when the initial guesses {$\bbeta_\TG$, $\bbeta_\rand$ and
  $\bbeta_\Hou$} are used. On the other hand, with the initial guess
{$\bbeta_\Kerr$}, a much lower (approximately by a factor of 4)
terminal value $\tPhi_{25}$ of the objective functional is found,
cf.~figure \ref{fig:init:25}(a), which corresponds to an optimal
initial condition with less energy in the high-wavenumber part of the
spectrum {in figure \ref{fig:init:25}(b)}.  The reason for this is
that the initial condition {$\bbeta_\Kerr$} is designed to produce a
large growth of different diagnostic quantities only after a long
time, approximately $t \approx 110059$ (this value of $t$ is obtained
by rescaling the time when \citet{HouLi2006} stopped their
computations to account for the normalization $\| \bbeta
\|_{\dot{H}^3} = 1$ we use in the present study). In the remainder of
this subsection we will use $\bbeta_\TG$ as the initial guess in
\eqref{eq:RCG}, which will allow us to compare our results with the
findings of earlier studies of potential singularities in Euler flows
that used the Taylor-Green vortex as the initial condition
\citep{bb12}.

\begin{figure}
  \centering
\mbox{\subfigure[]{\includegraphics[width=0.48\textwidth]{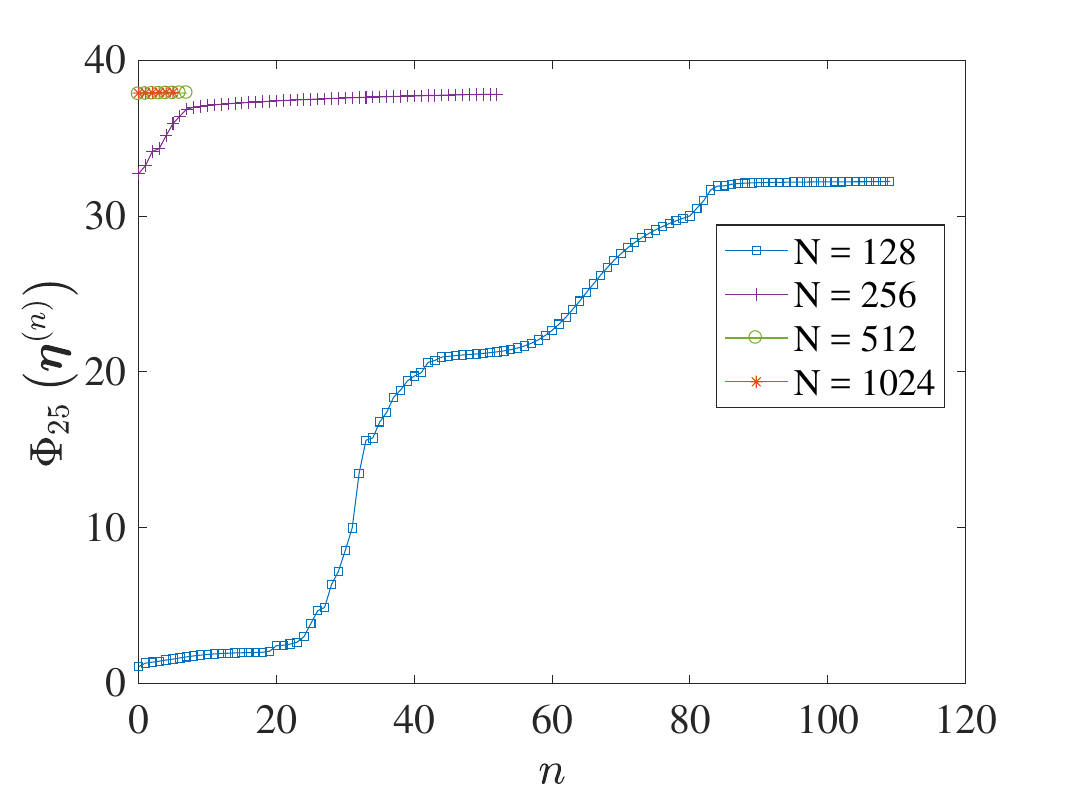}}
\hfill
\subfigure[]{\includegraphics[width=0.48\textwidth]{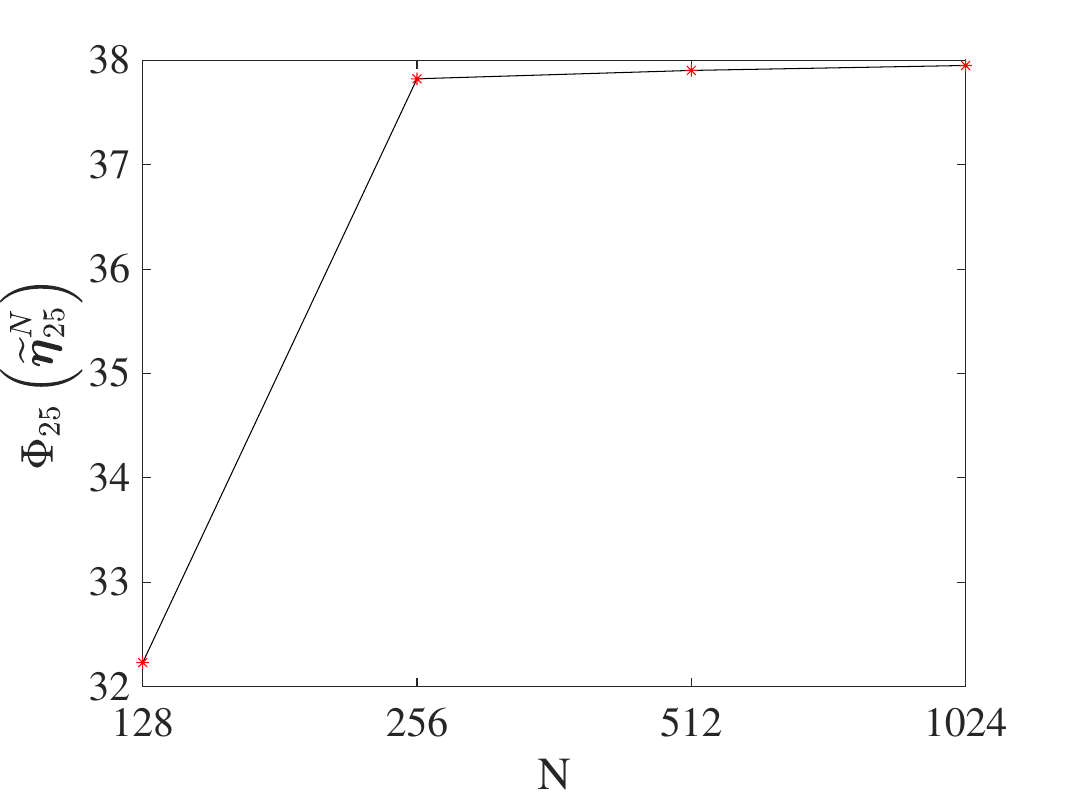}}}
\caption{[Short time window, $T = 25$] Dependence of (a) the objective
  functional $\Phi_{25}^N\left(\bbeta^{(n)}\right)$ on the iteration
  index $n$ for different resolutions $N^3$ and (b) of the
  corresponding maximum attained values $\tPhi_{25}^N$ of
  the objective functional on $N$.}
  \label{fig:resol:iter:25}
\end{figure}

As is evident from figure \ref{fig:init:25}(b), calculations performed
with the resolution $128^3$ are not fully resolved since in all cases
the optimal initial conditions $\teta_{25}$ have Fourier coefficients
with magnitudes larger than the machine precision for wavenumbers
$|\bk| > k_0 {:= 2\pi\lfloor N/3\rfloor}$, which results in
aliasing errors. To address this issue, we solve Problem~\ref{pb:H3}
with increasing resolutions $N^3 = 256^3, 512^3, 1024^3$, which is
done using the optimal initial condition $\teta_{25}^N$ as the initial
guess for the iteration~\eqref{eq:RCG} performed with the resolution
$(2N)^3$.  The dependence of the objective functional
$\Phi_{25}^N\left(\bbeta^{(n)}\right)$ on the iteration index $n$ is
shown in figure \ref{fig:resol:iter:25}(a), where we see that larger
values of the objective functional are achieved each time the
resolution is refined. We also observe that after each resolution
refinement, fewer iterations are needed to achieve convergence; this
is because only the Fourier coefficients with high wavenumbers, which
represent the fine structures in the flow, need to be adjusted
following a resolution refinement. As is evident from figure
\ref{fig:resol:iter:25}(b), the maximum values of the objective
functional $\tPhi_{25}^N$ converge to a well-defined finite limit as
the resolution increases, cf.~\eqref{eq:wellposed}.

\begin{figure}
  \centering
\mbox{\subfigure[]{\includegraphics[width=0.48\textwidth]{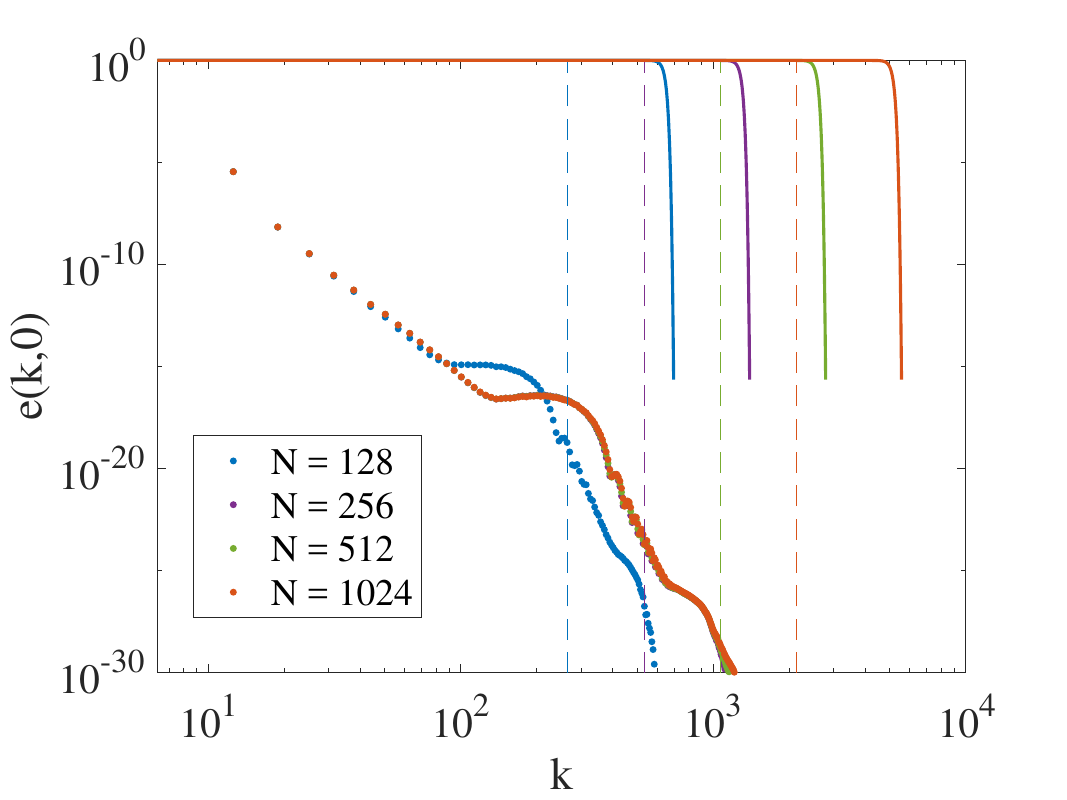}}
\hfill
\subfigure[]{\includegraphics[width=0.48\textwidth]{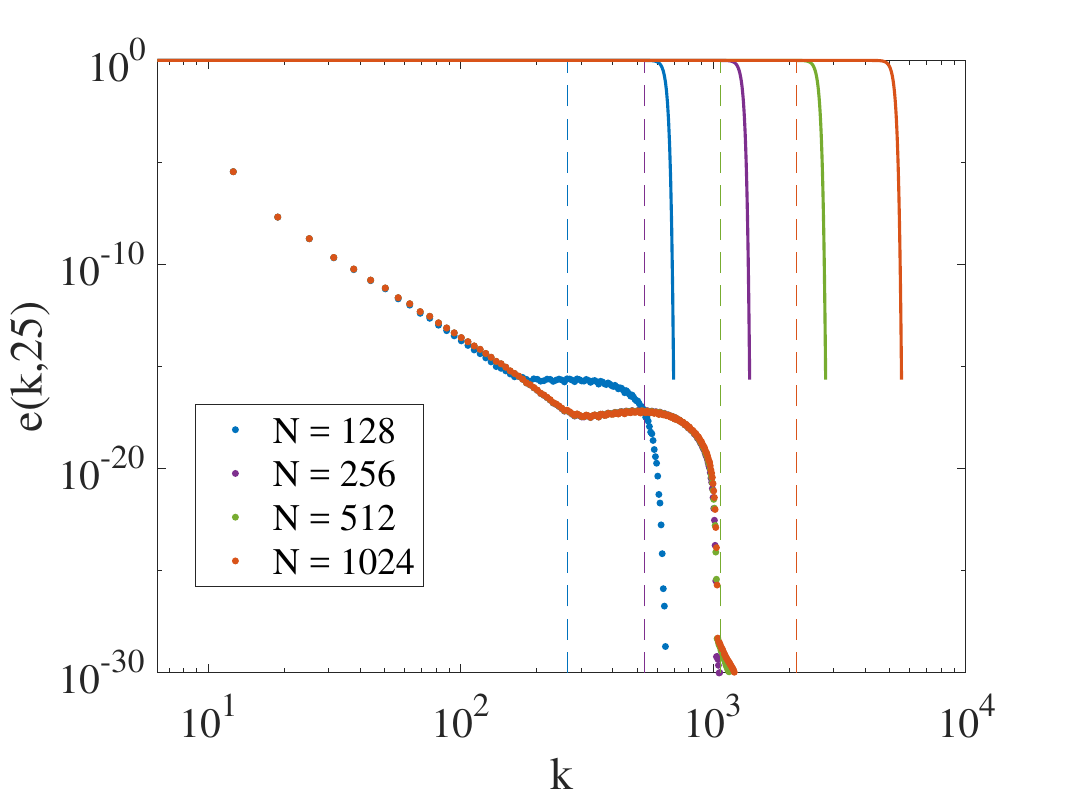}}}
  \caption{[Short time window, $T = 25$] The energy spectra of (a) the
    optimal initial conditions $\teta_{25}^N$ and (b) of the
    corresponding terminal states $\u^N\left(25; \teta_{25}^N\right)$
    obtained for different resolutions $N^3$. The solid lines
    represent the Gaussian filters we use \citep{hl07} whereas the
    dashed lines mark the threshold wavenumber $k_0$ above which
    aliasing errors occur.}
  \label{fig:spectrum:resol:25}
\end{figure}

\begin{figure}
  \centering
\mbox{\subfigure[]{\includegraphics[width=0.48\textwidth]{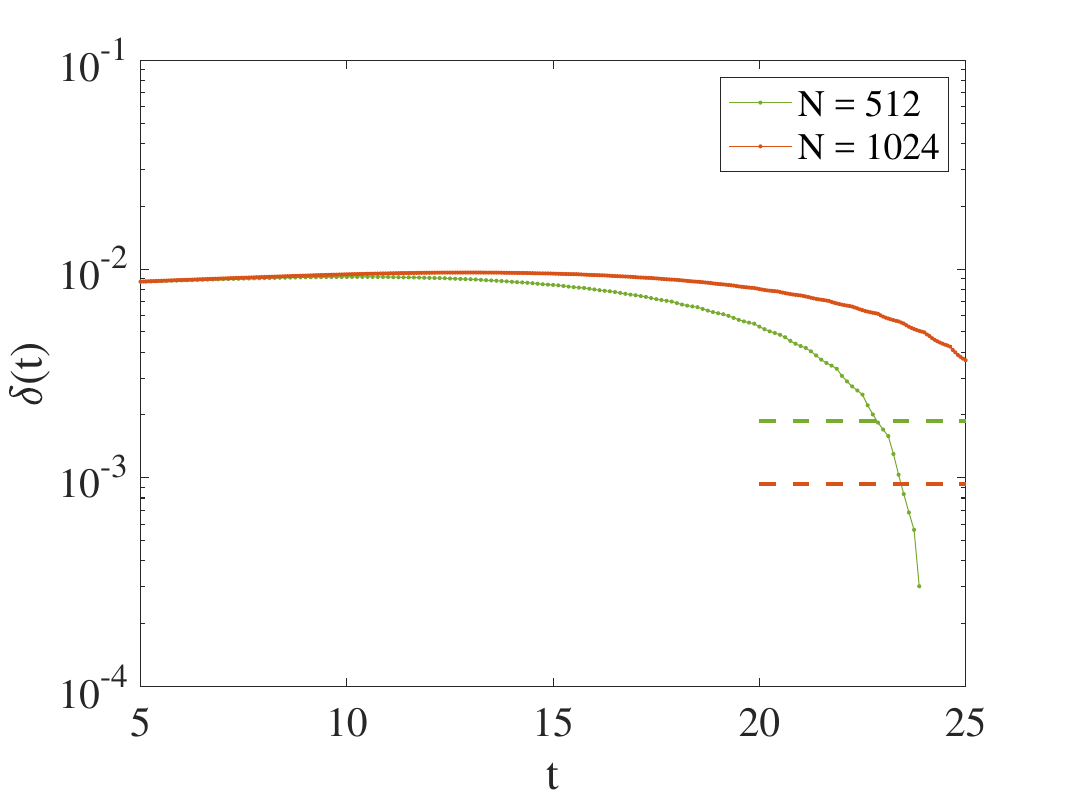}}
\hfill
\subfigure[]{\includegraphics[width=0.48\textwidth]{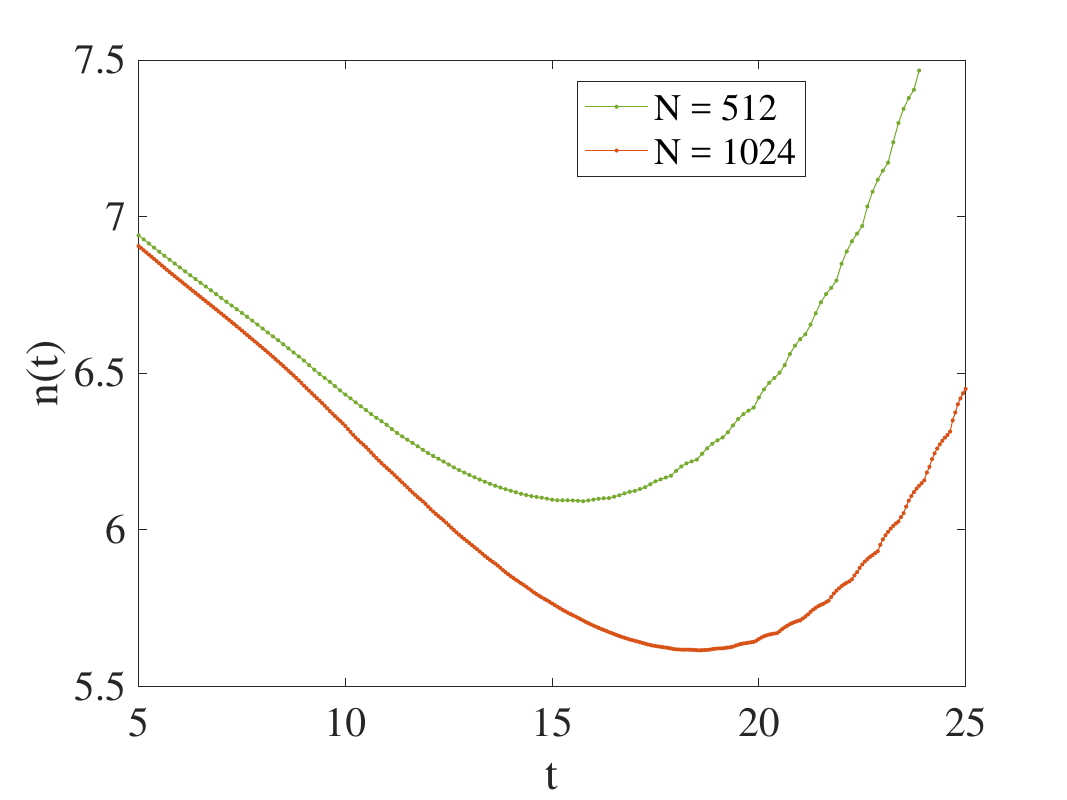}}}
\caption{[Short time window, $T = 25$] Dependence of (a) the width
  $\delta(t)$ of the analyticity strip and (b) the corresponding order
  $n(t)$ of the singularity, cf.~\eqref{eq:Ekinf}, on time $t \in
  [0,25]$ in the extreme flows computed with resolutions $N^3=512^3,
  1024^3$. {In (a) the horizontal lines represent the reliability
    condition \eqref{eq:Trel} corresponding to different
    resolutions.}}
  \label{fig:strip:25}
\end{figure}

The energy spectra of the optimal initial conditions $\teta_{25}^N$
and of the corresponding terminal states $\u^N\left(25;
  \teta_{25}^N\right)$ are shown in figures
\ref{fig:spectrum:resol:25}(a) and \ref{fig:spectrum:resol:25}(b) for
different resolutions.  {These plots indicate that the computations are
well resolved with $N^3=1024^3$}.  The time evolution of the width
$\delta(t)$ of the analyticity strip and of the corresponding order
$n(t)$ of the singularity are shown for $N^3=512^3, 1024^3$ in figures
\ref{fig:strip:25}(a) and \ref{fig:strip:25}(b), respectively.
{When determining these parameters by minimizing \eqref{eq:chi} we
  use $k_> = \min \{k_0, \ \sup_{e(k,t) \geq 10^{-30}} k \}$, i.e.,
  the spectrum was fitted up to the maximum wavenumbers unaffected by
  aliasing or to wavenumbers at which the magnitude of the Fourier
  coefficients would drop to the level of the machine precision. The
  fitting was also terminated whenever $\delta(t) \leq 0$.}
{Besides aliasing errors, we use another criterion to gauge the
  reliability of our computation which is based on $\delta(t)$.
  Following \citep{bmonmu83, bb12}, we define a ``reliability time''
  $\Trel$ using the condition
\begin{equation}
\label{eq:Trel}
\delta(\Trel)\, k_0 = 2
\end{equation}
and declare the numerical computation trustworthy for times $t \leq
\Trel$, or in other words, as long as $\delta(t) \geq 2/k_0$. In
figure \ref{fig:strip:25}(a) the reliability conditions
\eqref{eq:Trel} for different resolutions are marked with horizontal
lines. The reliability time increases as we increase the resolution
and the figure shows that the simulation remains reliable throughout
the entire time interval $[0, 25]$ when $N = 1024$, which is
consistent with figure \ref{fig:spectrum:resol:25}(b).  In figure
\ref{fig:strip:25}(a), $\delta(t)$ reveals a mild decrease only near
the end of the time window $[0,25]$.} These observations further
confirm that the extreme Euler flows which correspond to the initial
data $\teta_{25}$ obtained as local maximizers of Problem \ref{pb:H3}
for $T = 25$ remain regular on the interval $[0, 25]$.


\begin{figure}
  \centering
\mbox{\subfigure[]{\includegraphics[width=0.48\textwidth]{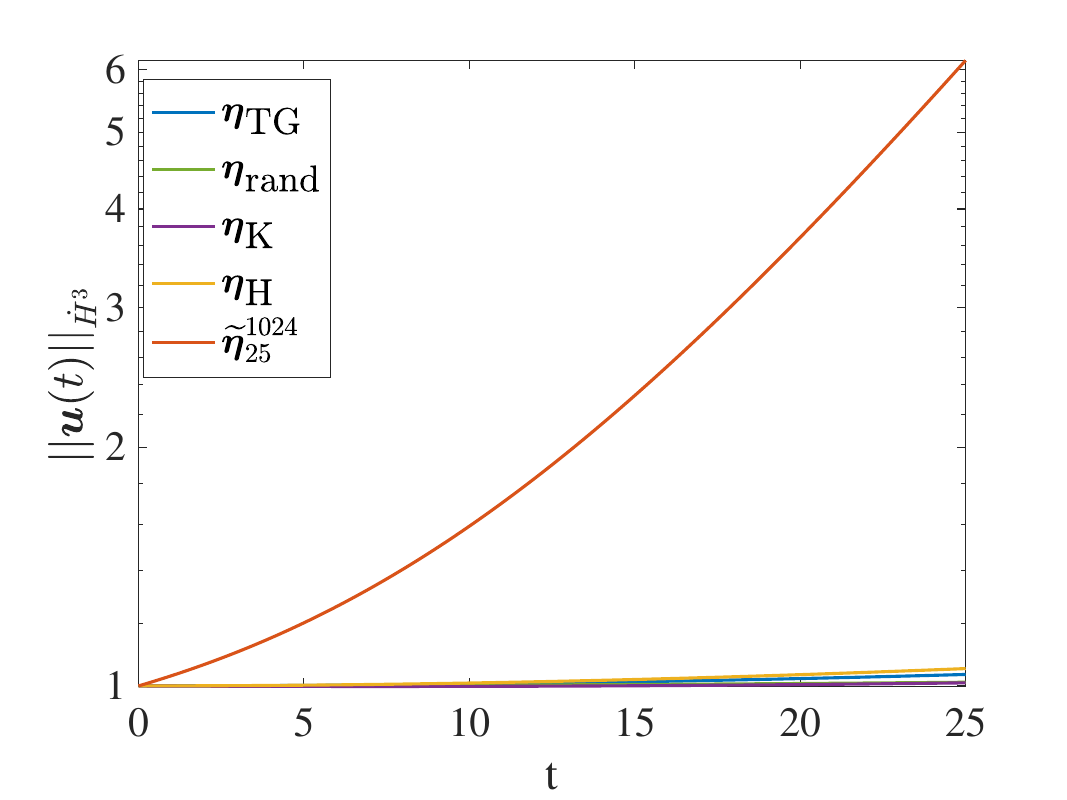}}
\hfill
\subfigure[]{\includegraphics[width=0.48\textwidth]{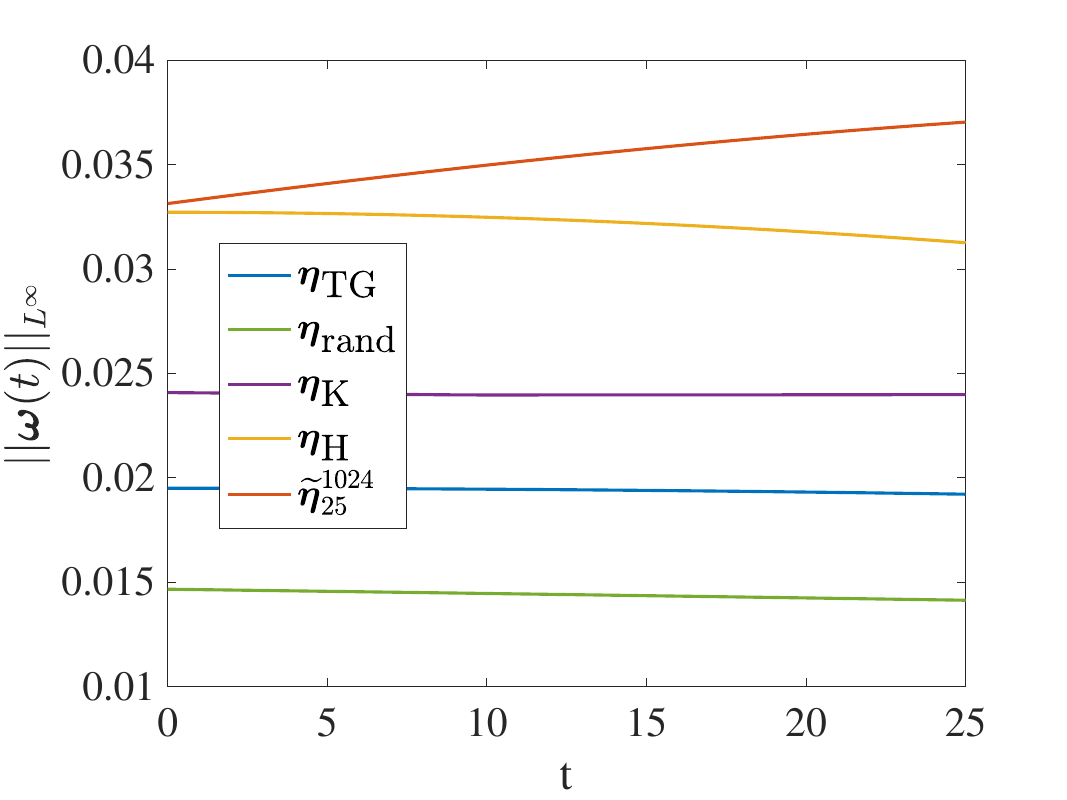}}}
  \caption{[Short time window, $T = 25$] Dependence of the solution
    norms (a) $\| \u(t) \|_{\dot{H}^3}$ and (b) $\| \bomega(t)
    \|_{L^\infty}$ for $t \in [0,25]$ in Euler flows corresponding to
    different initial conditions and approximated using the resolution
    $N^3 = 1024^3$.}
     \label{fig:norms:25}
   \end{figure}

To close this subsection, in figures \ref{fig:norms:25}(a) and
\ref{fig:norms:25}(b) we compare time evolution of the norms $\| \u(t)
\|_{\dot{H}^3}$ and $\| \bomega(t) \|_{L^\infty}$ in Euler flows
corresponding to different initial conditions listed in Section ~\ref{sec:IG}. 
We see that the growth of $\| \u(t) \|_{\dot{H}^3}$ is
by far the largest in the flow with the optimal initial condition
$\teta_{25}^{1024}$ and is only modest when the other initial
conditions are used. Likewise, the norm $\| \bomega(t) \|_{L^\infty}$
exhibits some weak growth only in the flow with the optimal initial
condition $\teta_{25}^{1024}$ while it actually decreases for $t \in
[0,25]$ when the initial conditions $\bbeta_{\TG}$, $\bbeta_{\rand}$
and $\bbeta_\Hou$ are used.

 
 %
   %
   
\subsection{Results for $T = 75$} 
\label{sec:Tlong}

\begin{figure}
  \centering
\mbox{\subfigure[]{\includegraphics[width=0.48\textwidth]{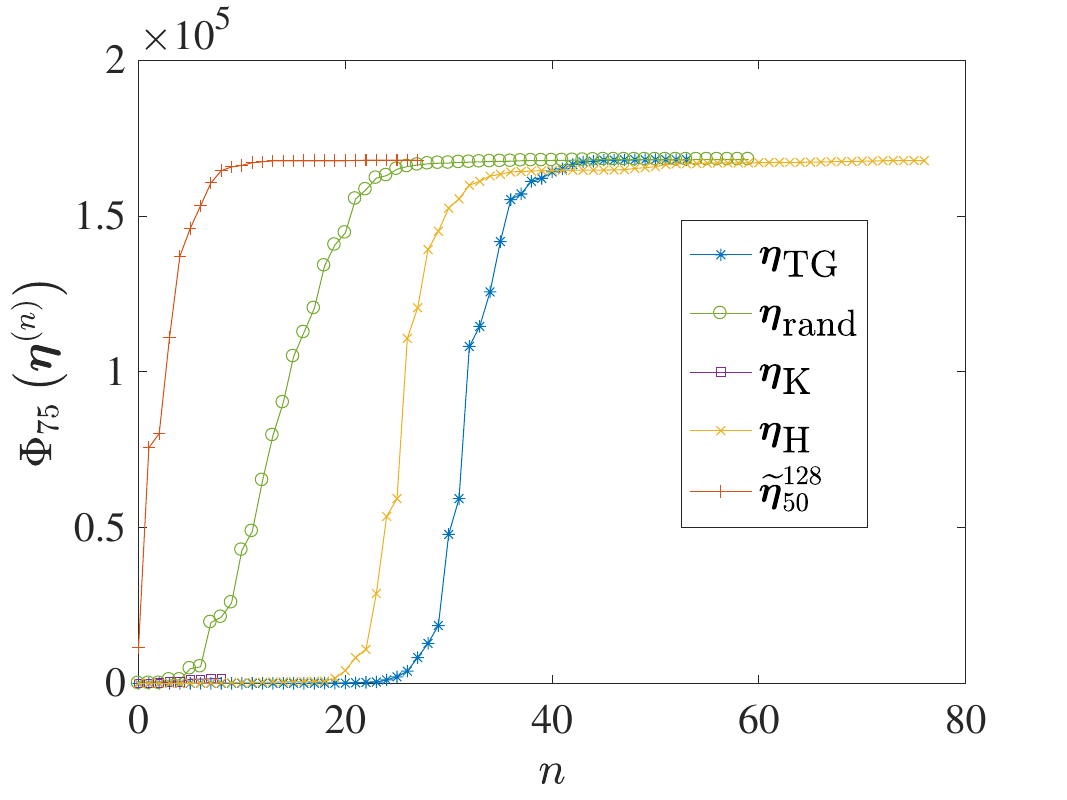}}
\hfill
\subfigure[]{\includegraphics[width=0.48\textwidth]{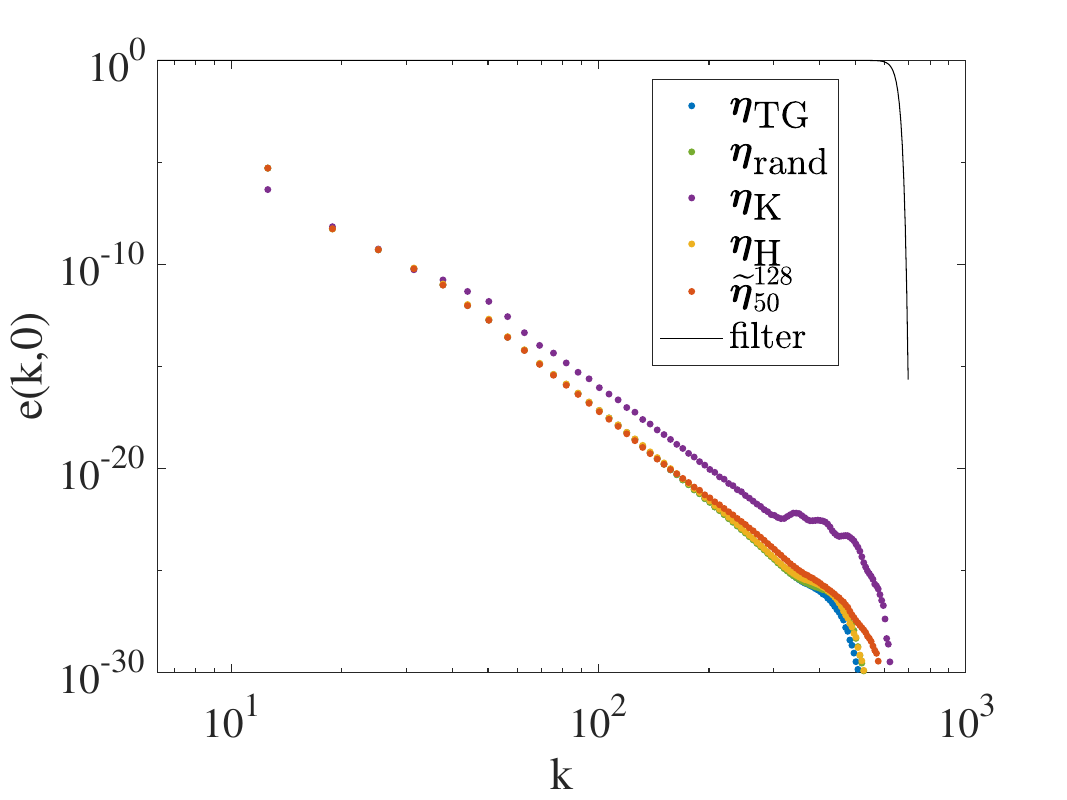}}}
  \caption{[Long time window, $T = 75$] (a) Dependence of the
    objective functional $\Phi_{75}\left(\bbeta^{(n)}\right)$ on the iteration
    index $n$ and (b) energy spectra \eqref{eq:Ek} of the optimal
    initial conditions $\teta_{75}$ corresponding to different initial
    guesses $\bbeta_{\TG}$, $\bbeta_{\rand}$,
    $\bbeta_{\Kerr}$ and $\teta_{25}$ used in iterations
    \eqref{eq:RCG}. In panel (b) the solid line represents the
    Gaussian filter we use \citep{hl07}.}
  \label{fig:init:75}
\end{figure}

We now move on to discuss solutions of Problem \ref{pb:H3} on a longer
time window with $T = 75$ where the behavior of the ``extreme'' flows
is qualitatively different from their behavior on the short time
window discussed in Section \ref{sec:Tshort}. We begin our discussion by
analyzing the effect of the initial guess $\bbeta_0$ on the local
maximizers of Problem \ref{pb:H3} obtained with the iterative
algorithm \eqref{eq:RCG}.  In figure \ref{fig:init:75}(a) we show the
dependence of the objective functional
$\Phi_{75}\left(\bbeta^{(n)}\right)$ on the iteration index $n$ and
figure \ref{fig:init:75}(b) shows the energy spectra of the
corresponding optimal initial conditions $\teta_{75}$.  We remark
that, as an intermediate step, we first use the optimal initial
condition $\teta_{25}^{128}$ as the initial guess to solve Problem
\ref{pb:H3} with $T = 50$, and then use its solution,
$\teta_{50}^{128}$, as the initial guess for $T = 75$.  We see that,
with the exception of the initial guess $\bbeta_\Kerr$, computations
performed with all other initial guesses lead to the same local
maximizer (up to rotation and translation). As was the case when
Problem \ref{pb:H3} was solved with $T = 25$, computations using the
initial guess $\bbeta_\Kerr$ lead to a much smaller maximum
attained value of the objective functional since this initial
condition is designed to promote a significant growth of various
regularity indicators, such as $\| \u(t) \|_{\dot{H}^3}$ and $||\bds
\omega(t) ||_{L^\infty}$, only on much longer time scales
\citep{HouLi2006}.

\begin{figure}
    \centering
\mbox{\subfigure[]{\includegraphics[width=0.48\textwidth]{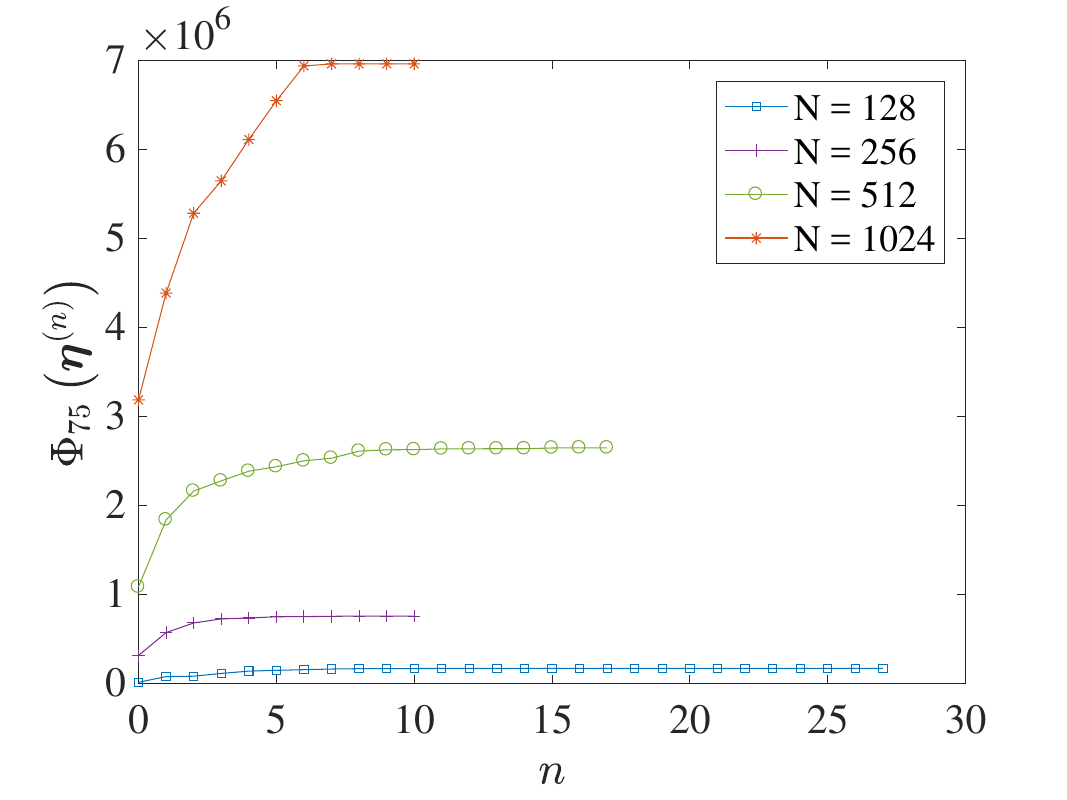}}
\hfill
\subfigure[]{\includegraphics[width=0.48\textwidth]{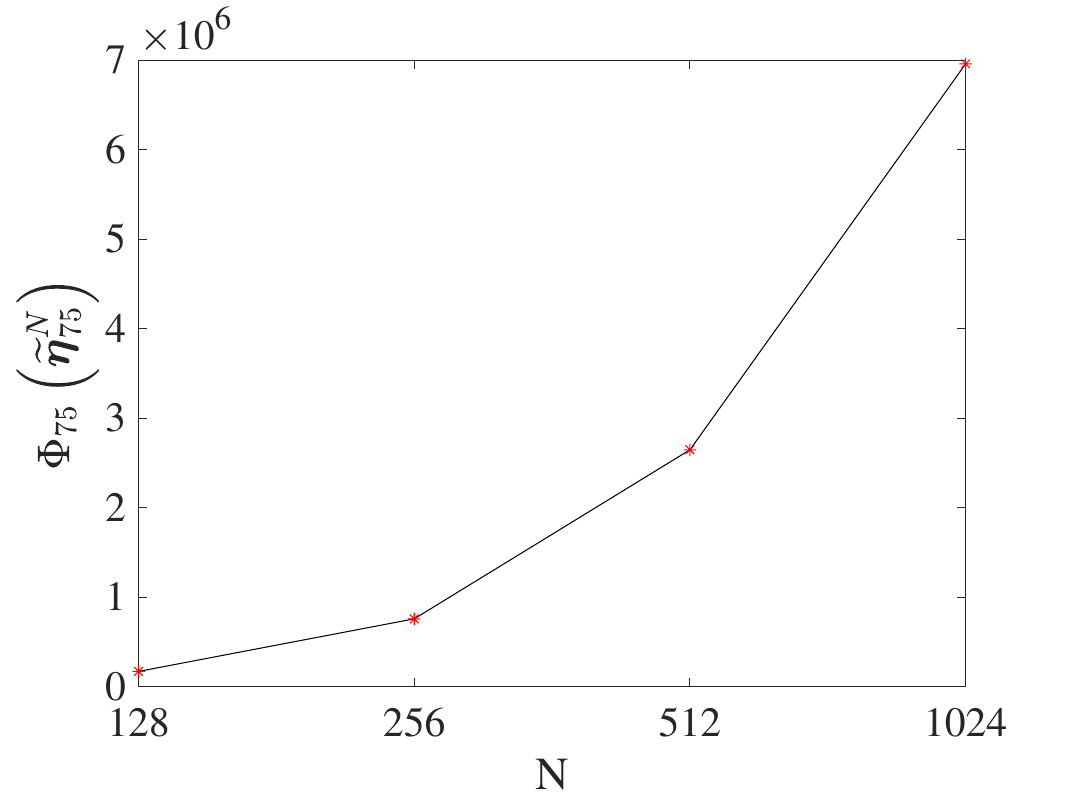}}}
    \caption{[Long time window, $T = 75$] Dependence of (a) the
      objective functional $\Phi_{75}^N\left(\bbeta^{(n)}\right)$ on
      the iteration index $n$ for different resolutions $N^3$ and (b)
      of the corresponding maximum attained values $\tPhi_{75}^N$ of
      the objective functional on $N$.}
  \label{fig:resol:iter:75}
  \end{figure}

As is evident from figure \ref{fig:init:75}(b), computations carried
out with the resolution $128^3$ are under-resolved and now we consider
the effect of refining the resolution on solutions of Problem
\ref{pb:H3} with $T = 75$. As we did in Section \ref{sec:Tshort}, we
proceed by using the optimal initial condition $\teta_{75}^N$ as the
initial guess in iteration \eqref{eq:RCG} when solving
Problem \ref{pb:H3} with the resolution $(2N)^3$.  The dependence of
the objective functional
$\Phi_{75}^N\left(\bbeta^{(n)}\right)$ on the iteration
index $n$ is presented for different resolutions in
figure~\ref{fig:resol:iter:75}(a). Figure
  \ref{fig:resol:iter:75}(b) shows that, unlike in the case with $T =
  25$, cf.~figure \ref{fig:resol:iter:25}(b), the maximum attained
  values of the objective functional $\tPhi_{75}^N$ diverge as we
  refine the resolution with the difference $\tPhi_{75}^{2N} -
  \tPhi_{75}^N$ increasing with $N$. As explained in
Section \ref{sec:results}, this indicates the possibility of a
  singularity formation at some $t \in [0, 75]$.

\begin{figure}
    \centering
\mbox{\subfigure[]{\includegraphics[width=0.48\textwidth]{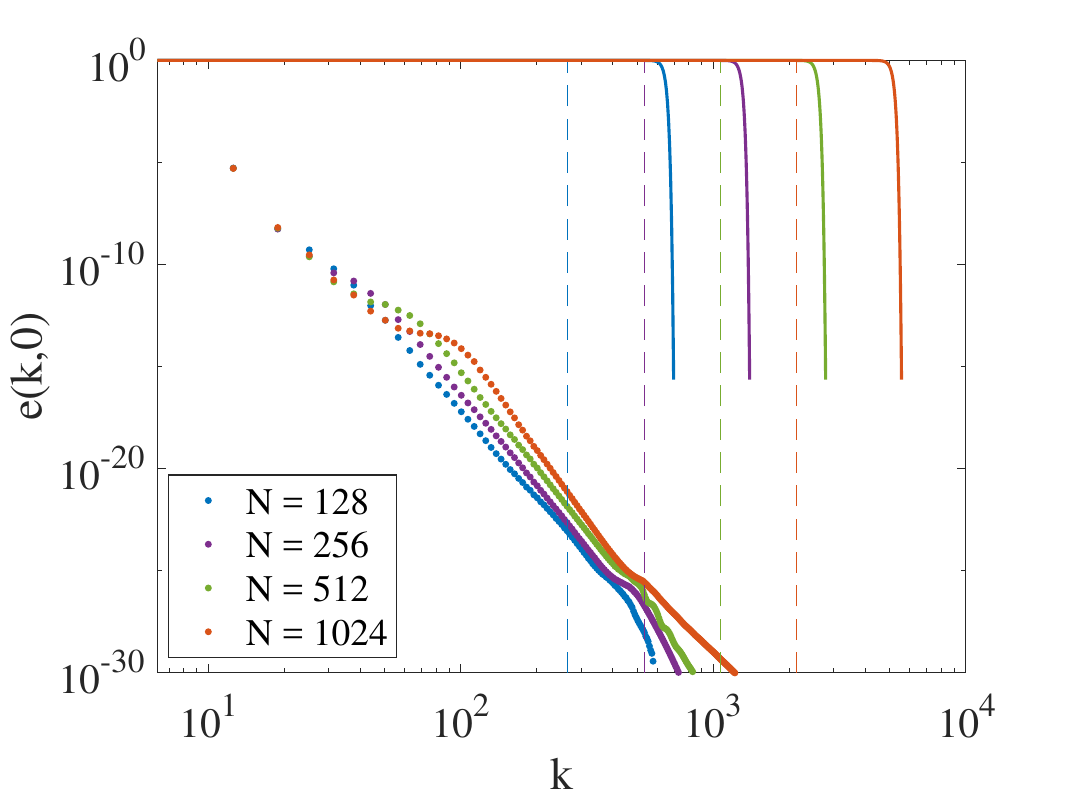}}
\hfill
\subfigure[]{\includegraphics[width=0.48\textwidth]{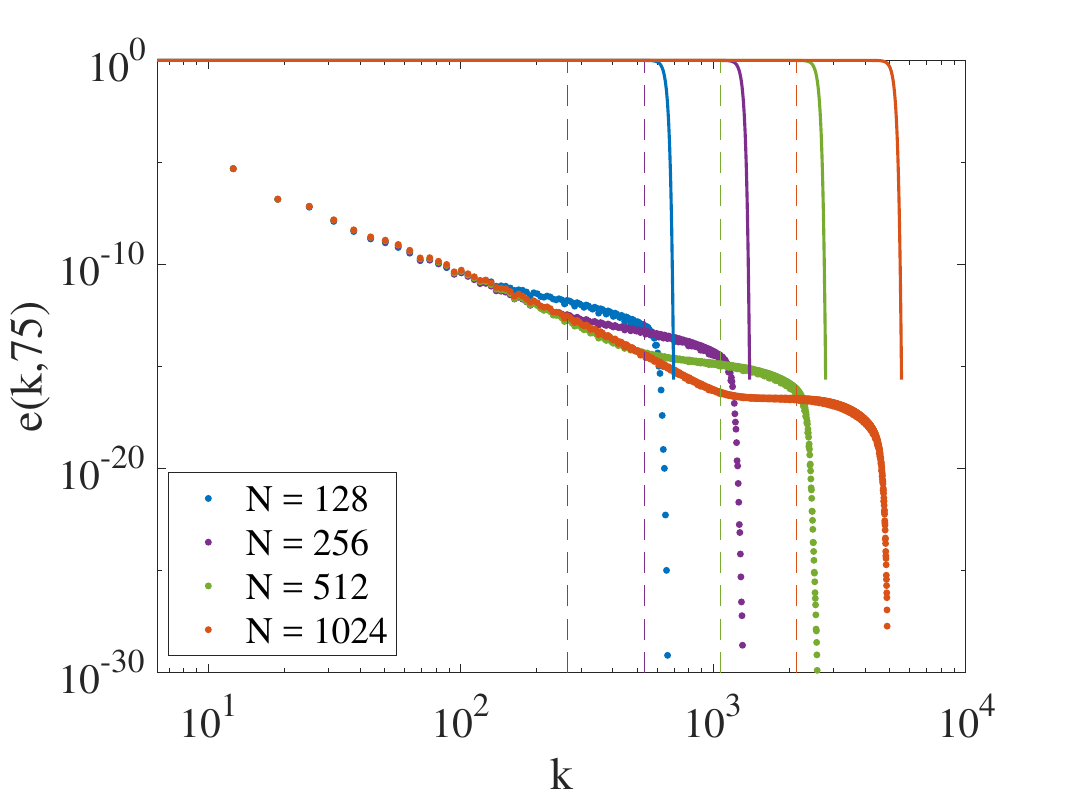}}}
    \caption{[Long time window, $T = 75$] The energy spectra of (a)
      the optimal initial conditions $\teta_{75}^N$ and (b) of the
      corresponding terminal states $\u^N\left(75;
        \teta_{75}^N\right)$ obtained for different resolutions $N^3$.
      The solid lines represent the Gaussian filters we use
      \citep{hl07} whereas the dashed lines mark the threshold
      wavenumber $k_0$ above which aliasing errors occur.}
    \label{fig:spectrum:resol:75}
  \end{figure}

  The energy spectra of the optimal initial conditions $\teta_{75}^N$
  and the corresponding terminal states $\u^N\left(75;
    \teta_{75}^N\right)$ are shown in figures
  \ref{fig:spectrum:resol:75}(a) and \ref{fig:spectrum:resol:75}(b),
  respectively. Comparing these two figures, we observe that even
  though the initial conditions $\teta_{75}^N$ are well-resolved for
  $N = 512, 1024$, the time evolutions computed with all resolutions
  become under-resolved at later times.  In particular, for $N = 1024$
  this happens at {$t = T_0:= 51.25$} and is detected by the
  appearance of aliasing errors when the amplitudes of Fourier
  coefficients with {wavenumbers $|\bk| > k_0$ become larger than
    $10^{-30}$.}  For lower resolutions, computations become
  under-resolved at earlier times. {Once the numerical solution
    becomes under-resolved, its blow-up is prevented by the presence
    of the filter $\widehat G_{\bds j}$ (cf.~Section
    \ref{sec:num:discretization}) which can be regarded as a form of
    dissipative regularization applied to the Euler system
    \eqref{eq:Euler}. Since the form of the filter depends on the
    resolution $N$ (cf.~figures \ref{fig:spectrum:resol:25} and
    \ref{fig:spectrum:resol:75}), its regularizing effect vanishes
    when the resolution is refined, thereby allowing the computed
    solution to approach the singular trajectory as seen in figure
    \ref{fig:resol:iter:75}(b).}

\begin{figure}
   \centering
\mbox{\subfigure[]{\includegraphics[width=0.48\textwidth]{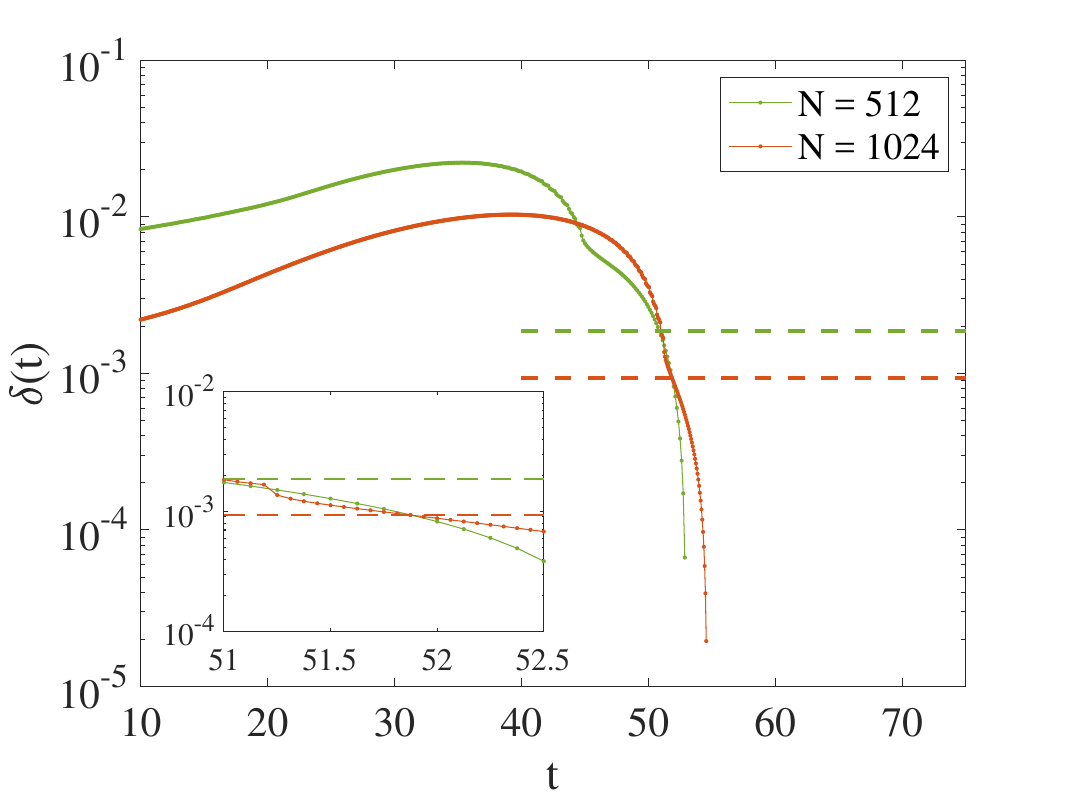}}
\hfill
\subfigure[]{\includegraphics[width=0.48\textwidth]{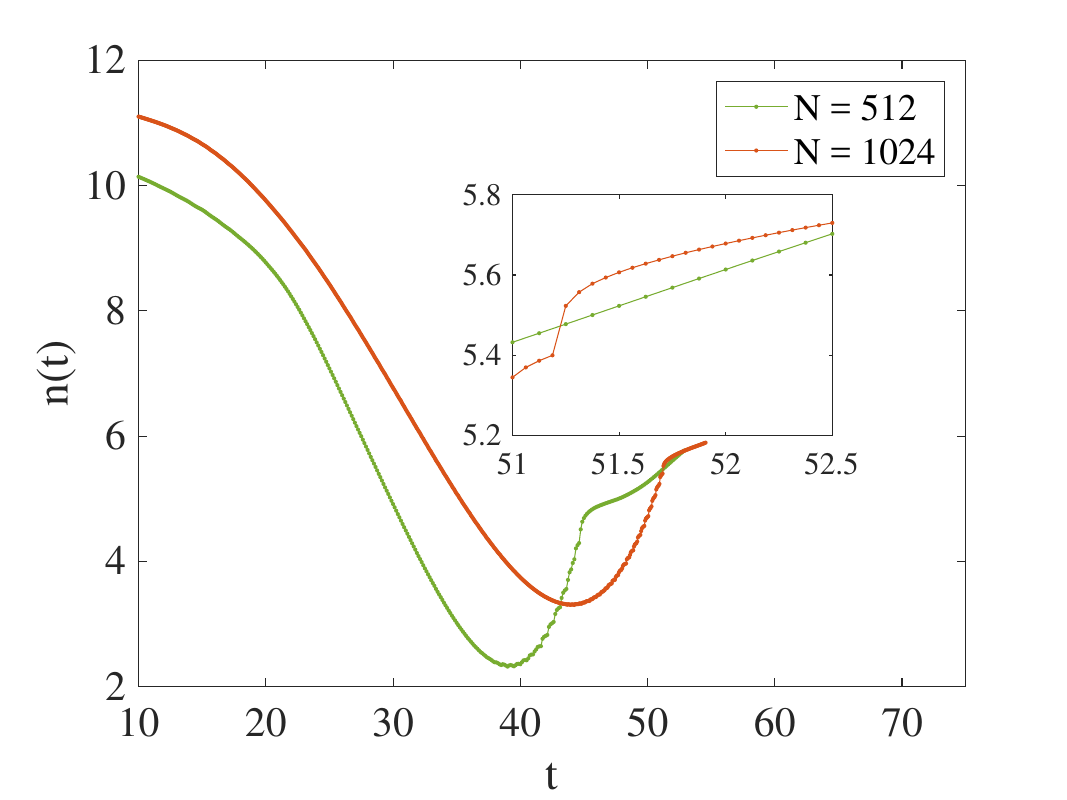}}}
\caption{[Long time window, $T = 75$] Dependence of (a) the width
  $\delta(t)$ of the analyticity strip and (b) the corresponding order
  $n(t)$ of the singularity, cf.~\eqref{eq:Ekinf}, on time $t \in
  [0,75]$ in the extreme flows computed with resolutions $N^3=512^3,
  1024^3$. {In (a) the horizontal lines represent the reliability
    condition \eqref{eq:Trel} corresponding to different
    resolutions.}}
  \label{fig:strip:75}
 \end{figure}

  The time evolution of the width $\delta(t)$ of the analyticity strip
  and of the corresponding order $n(t)$ of the singularity in optimal
  flows computed with $N=512,1024$ is shown in figures
  \ref{fig:strip:75}(a) and \ref{fig:strip:75}(b), respectively.
  {When determining these parameters by minimizing \eqref{eq:chi} we
    use $k_> = k_0$, i.e., the spectrum was fitted up to the maximum
    wavenumbers unaffected by aliasing. The fitting was also
    terminated whenever $\delta(t) \leq 0$.}  {In figure
    \ref{fig:strip:75}(a) we see that, following an initial period of
    growth, $\delta(t)$ starts to decrease when $t > 40$ and this
    decrease follows a power-law behavior.  According to the
    reliability conditions \eqref{eq:Trel} indicated in figure
    \ref{fig:strip:75}(a), when $N = 1024$, we can trust the numerical
    computation up to time $t = \Trel = 51.875$, which is close to,
    but slightly longer, than the time $T_0$ when aliasing errors appear.
    This is reasonable as there is a delay in the effect of aliasing
    errors on the part of the spectrum which determines the width of
    the analyticity strip.}


%

\begin{figure}
  \centering
\mbox{\subfigure[]{\includegraphics[width=0.48\textwidth]{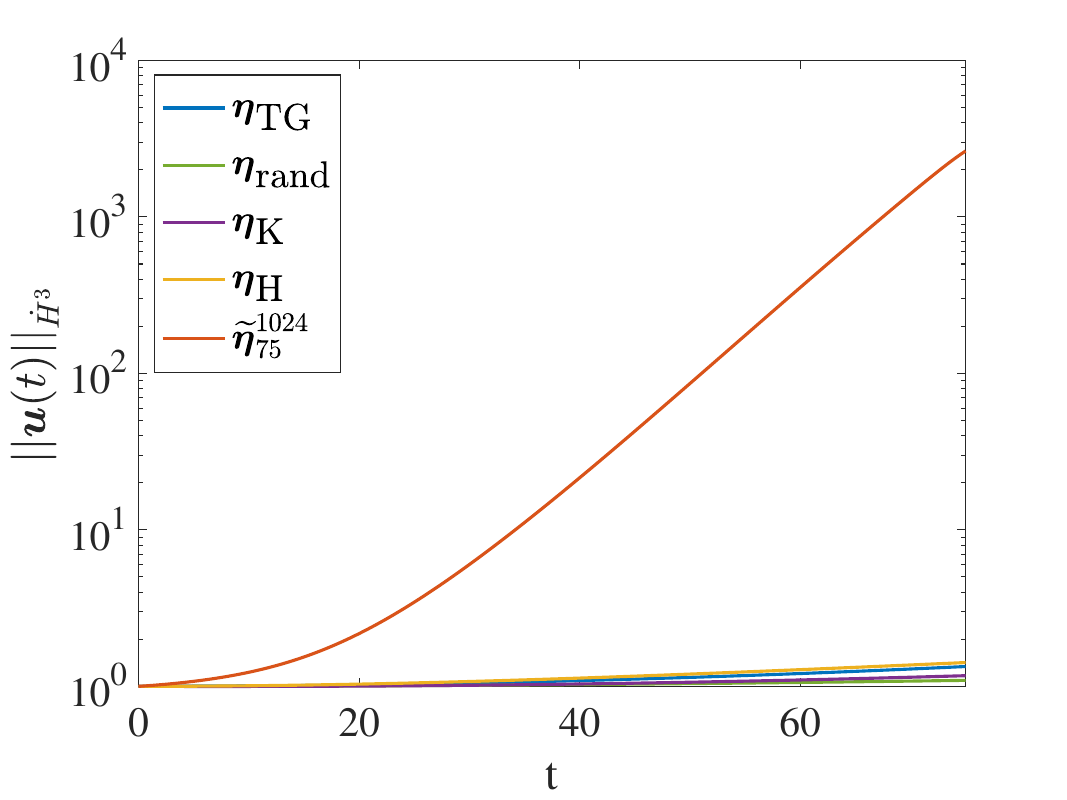}}
\hfill
\subfigure[]{\includegraphics[width=0.48\textwidth]{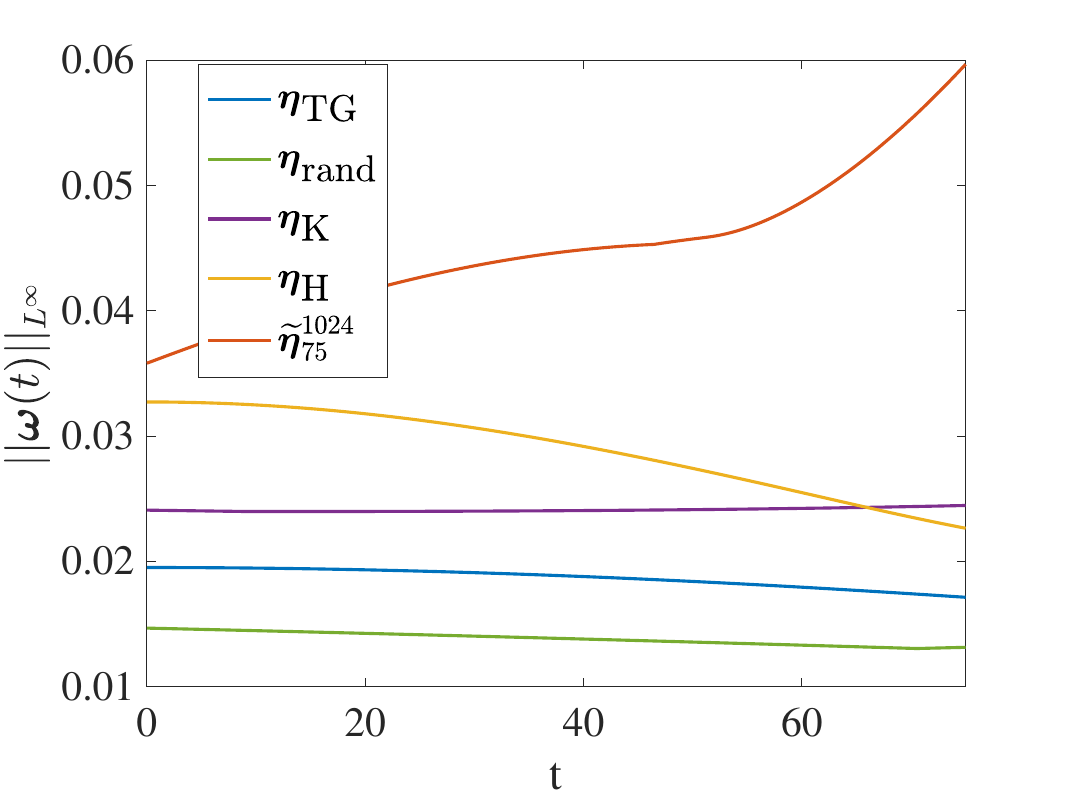}}}
  \caption{[Long time window, $T = 75$] Dependence of the solution
    norms (a) $\| \u(t) \|_{\dot{H}^3}$ and (b) $\| \bomega(t)
    \|_{L^\infty}$ for $t \in [0,75]$ in Euler flows corresponding to
    different initial conditions and approximated using the resolution
    $N^3 = 1024^3$.}
  \label{fig:H3:max:compare:75}
\end{figure}

 The time evolution of $\| \u(t) \|_{\dot{H}^3}$ and $\| \bomega(t)
 \|_{L^\infty}$ in the flows corresponding to different initial
 conditions is shown in figures \ref{fig:H3:max:compare:75}(a) and
 \ref{fig:H3:max:compare:75}(b), respectively. As was the case in
 Section \ref{sec:Tshort}, a significant growth of these norms is
 evident only in the flows corresponding to the optimal initial
 condition $\teta_{75}^{1024}$, whereas $\| \bomega(t) \|_{L^\infty}$
 actually decreases on $[0, 75]$ for solutions with the initial
 conditions $\bbeta_{\TG}$, $\bbeta_{\rand}$ and $\bbeta_{\Hou}$.
 {As regards the time evolution of the latter quantity,
   \citet[Corollaries 10 and 11]{bb12} used the BKM criterion
   \eqref{eq:BKM:vort} to obtain a condition (whose details are
   omitted for brevity) which must be satisfied by the exponent
   $\Gamma(t) > 0$ in an ansatz describing the evolution of the width
   of the analyticity strip under the assumption of finite-time
   blow-up, namely $\delta(t)= C(t)(T^* - t)^{\Gamma(t)}$, and the
   corresponding evolution of the order of singularity $n(t)$. The
   flow evolution corresponding to the optimal initial data
   $\teta_{75}^{1024}$ satisfies this condition, indicating that this
   evolution is consistent with a possible singularity formation in
   finite time despite modest growth of the norm $\| \bomega(t)
   \|_{L^\infty}$ evident in figure \ref{fig:H3:max:compare:75}(b).}

\begin{figure}
  \centering
\mbox{\subfigure[]{\includegraphics[width=0.48\textwidth]{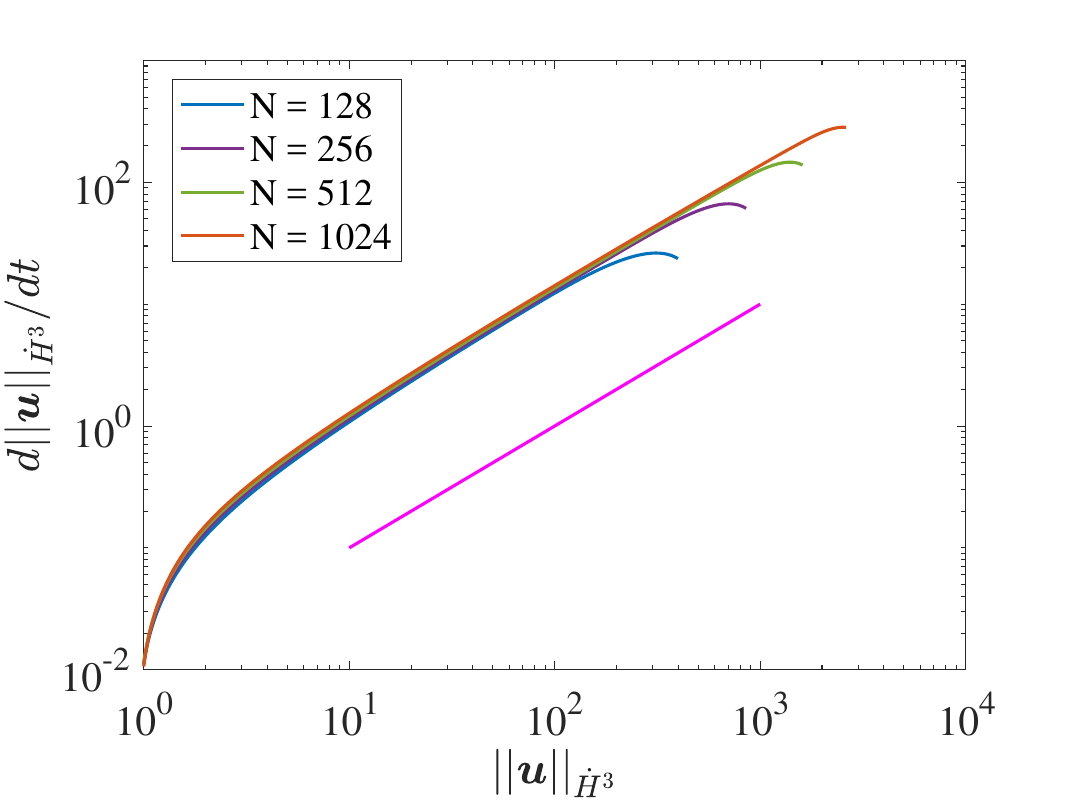}}
\hfill
\subfigure[]{\includegraphics[width=0.48\textwidth]{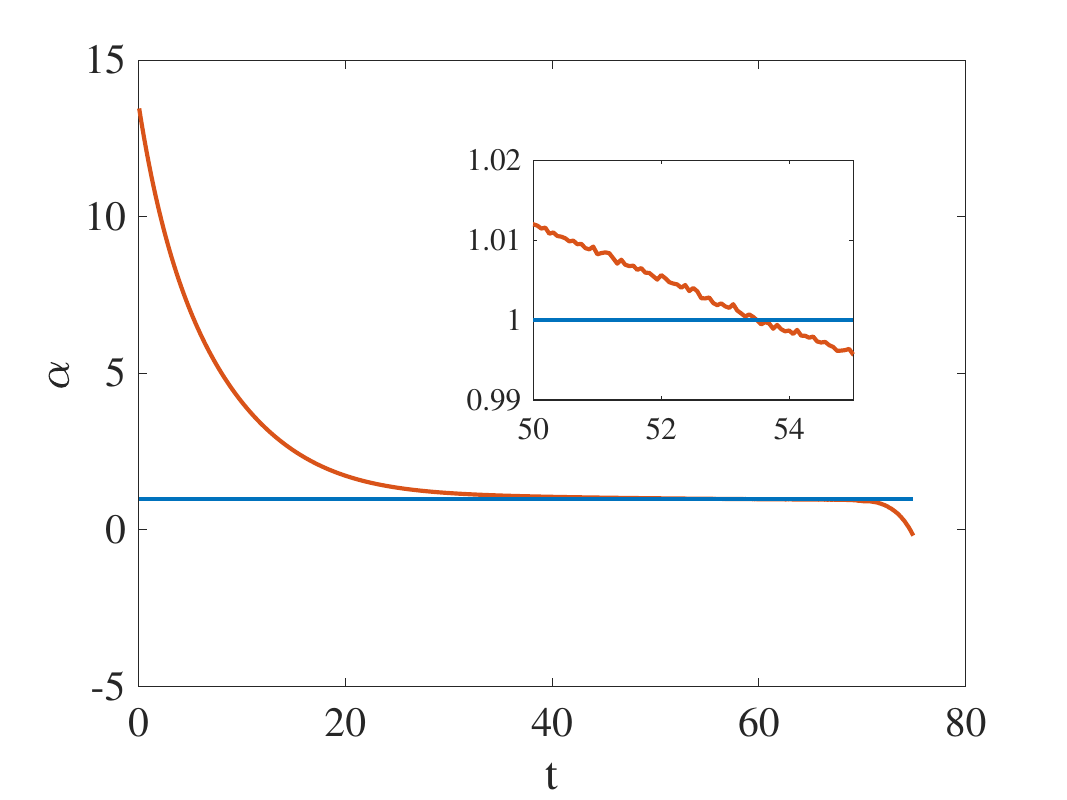}}}
  \caption{[Long time window, $T = 75$] (a) Dependence of $(d/dt)\|
    \u(t) \|_{\dot{H}^3}$ on $\| \u(t) \|_{\dot{H}^3}$ in Euler flows
    with the optimal initial conditions $\teta_{75}^N$ for $t \in
    [0,75]$ and different resolutions $N^3$ and (b) the corresponding
    exponent $\alpha(t)$ in \eqref{eq:duH3dt} for $t \in [0,75]$ with
    $N = 1024$. The straight lines in both panels represent $\alpha =
    1$.}
  \label{fig:u:H3:rate}
\end{figure}

In order to shed light on whether the growth of the norm $\|
  \u(t) \|_{\dot{H}^3}$ observed in
figure~\ref{fig:H3:max:compare:75}(a) corresponding to the optimal
initial conditions $\teta_{75}^N$ may indicate a singularity formation
at some time $t \in [0,75]$, we further analyze the growth rate of $\|
\u(t) \|_{\dot{H}^3}$. We assume that the evolution of this norm is
described by the relation
 \begin{equation}
   \label{eq:duH3dt}
   \frac{d \|\bds u(t)\|_{\dot H^3}}{dt} = C(t) \|\bds u(t)\|_{\dot H^3}^{\alpha(t)},
 \end{equation}
 which is motivated by the structure of rigorous a priori estimates
 for the rate of growth of Sobolev norms of solutions to the
 Navier-Stokes system \citep{dg95,ld08,ap16}. We note that if the
 exponent $\alpha(t)$ in \eqref{eq:duH3dt} remains larger than 1 over
 a sufficiently long time with the prefactor $C(t)$ bounded away from
 zero, then there will be a finite-time blow-up of the norm $\| \u(t)
 \|_{\dot{H}^3}$, and thus formation of a singularity in the
 corresponding Euler flow, cf.~Theorem \ref{thm:Hm}. To obtain further
 insights about these quantities, in figure \ref{fig:u:H3:rate}(a) we
 plot $(d/dt)\| \u(t) \|_{\dot{H}^3}$ versus $\| \u(t) \|_{\dot{H}^3}$
 with $t \in [0, 75]$ using log-log scaling for the optimal flows
 computed with different resolutions, such that the exponent
 $\alpha(t)$ can be inferred from the slope of the tangent to the
 curves at $\| \u(t) \|_{\dot{H}^3}$. The evolution of the exponent
 $\alpha(t)$ determined by a local fitting procedure applied to ansatz
 \eqref{eq:duH3dt} with time $t \in [0,75]$ is shown for $N = 1024$ in
 figure \ref{fig:u:H3:rate}(b), where a decreasing trend is evident
 and we have $\alpha(t) > 1$ for $t \in [0, 53.495]$ while the
 computation becomes under-resolved at $t \approx 51.25$.  As regards
 the prefactor $C(t)$ in \eqref{eq:duH3dt}, it reveals a slow growth
 with time $t$ (and with $\| \u(t) \|_{\dot{H}^3}$) which is well
 approximated by the expression $C(t) = 0.0568 \left( \ln \| \u(t)
   \|_{\dot{H}^3} \right)^{0.5742}$ with parameters determined via a
 least-squares fit. We thus conclude that the time evolution of the
 norm $\| \u(t) \|_{\dot{H}^3}$ in the flow with the optimal initial
 condition $\teta_{75}^{1024}$ remains consistent with the singularity
 formation as long as the computation remains well-resolved.

\begin{figure}
  \centering
\mbox{\subfigure[]{\includegraphics[width=0.48\textwidth]{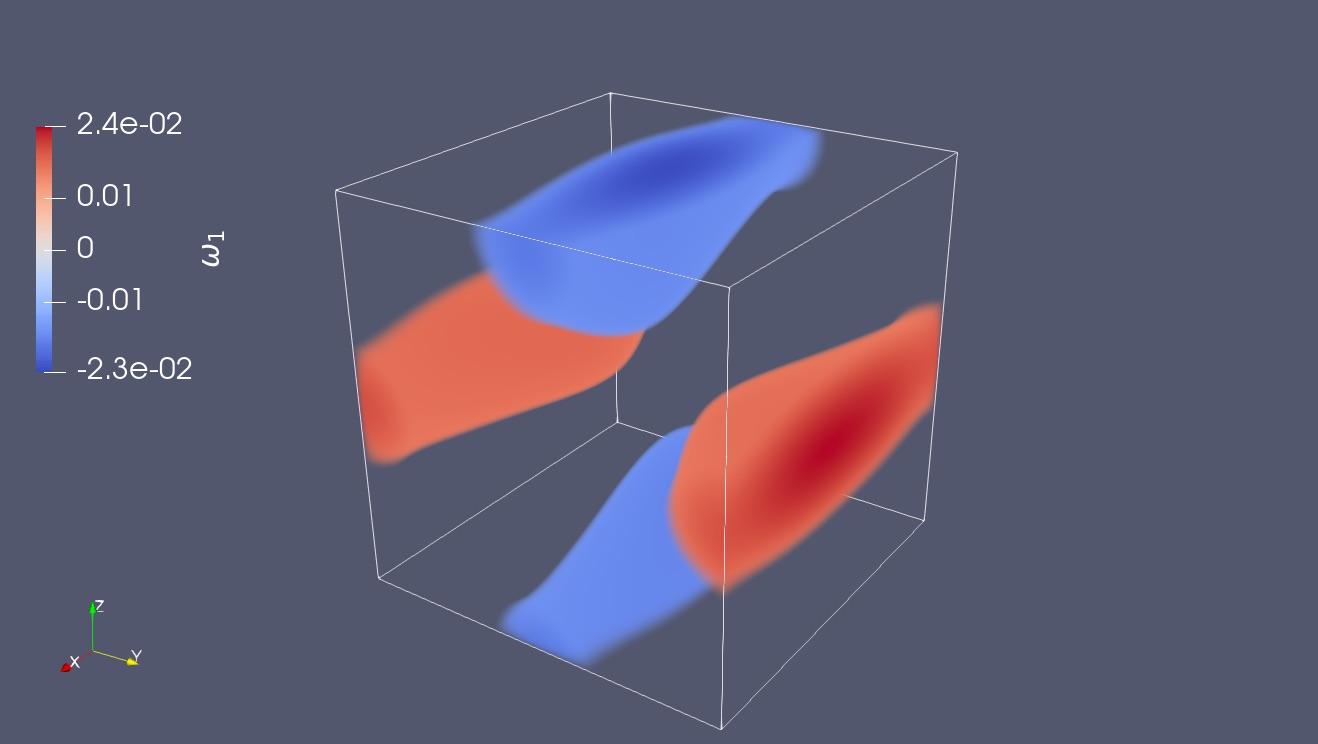}}
\hfill
\subfigure[]{\includegraphics[width=0.48\textwidth]{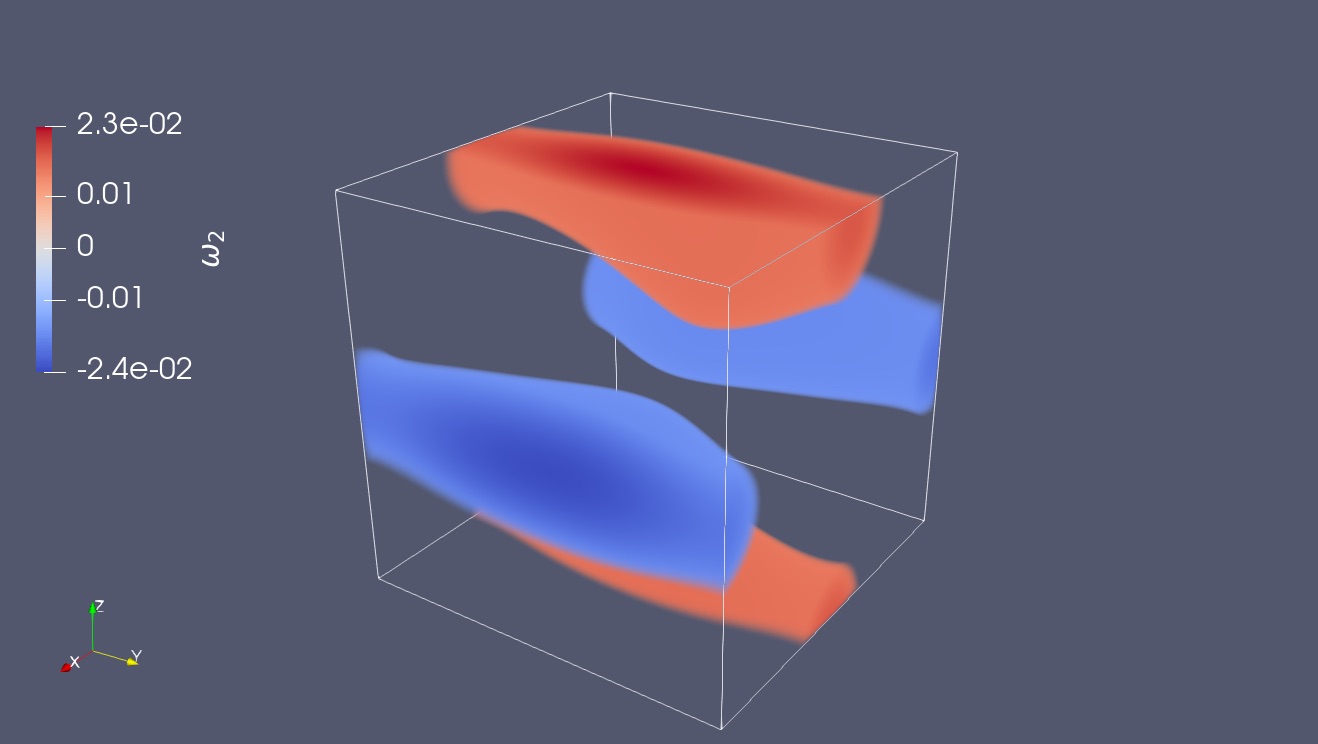}}}
\mbox{
\subfigure[]{\includegraphics[width=0.48\textwidth]{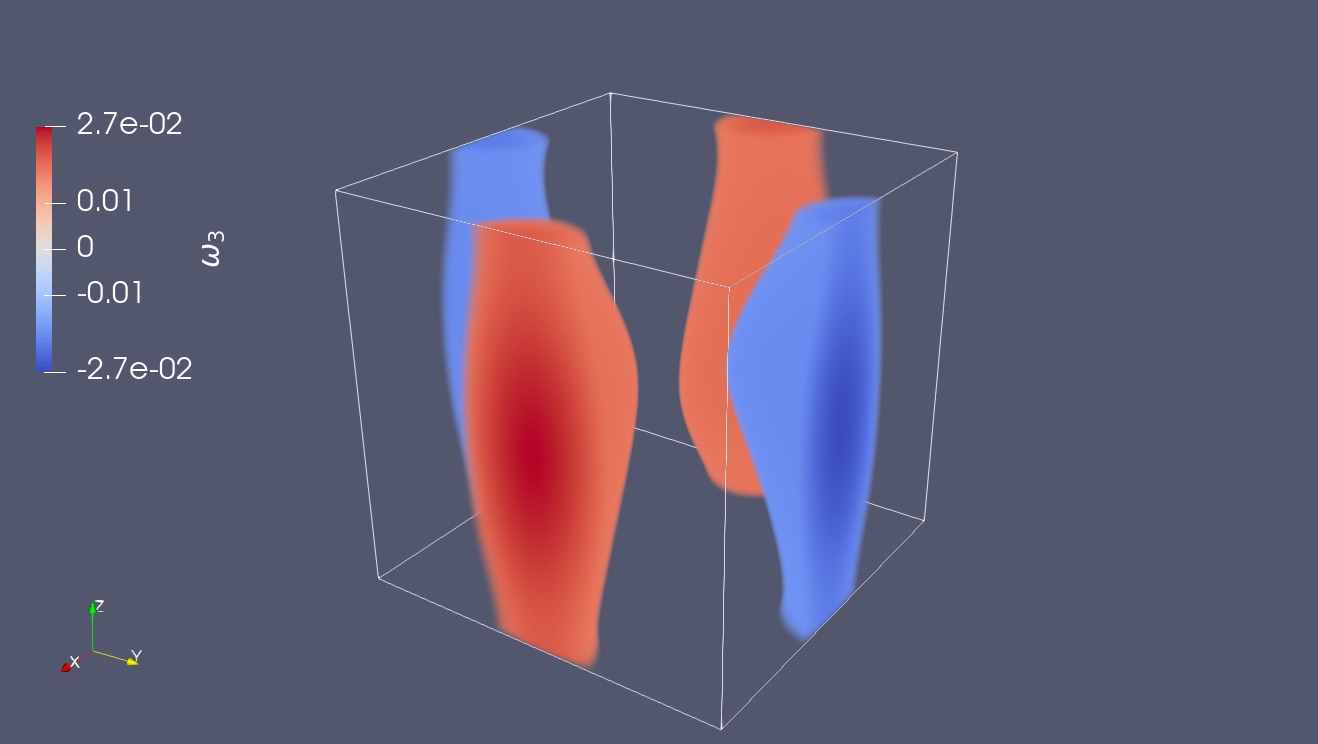}}
\hfill
\subfigure[]{\includegraphics[width=0.48\textwidth]{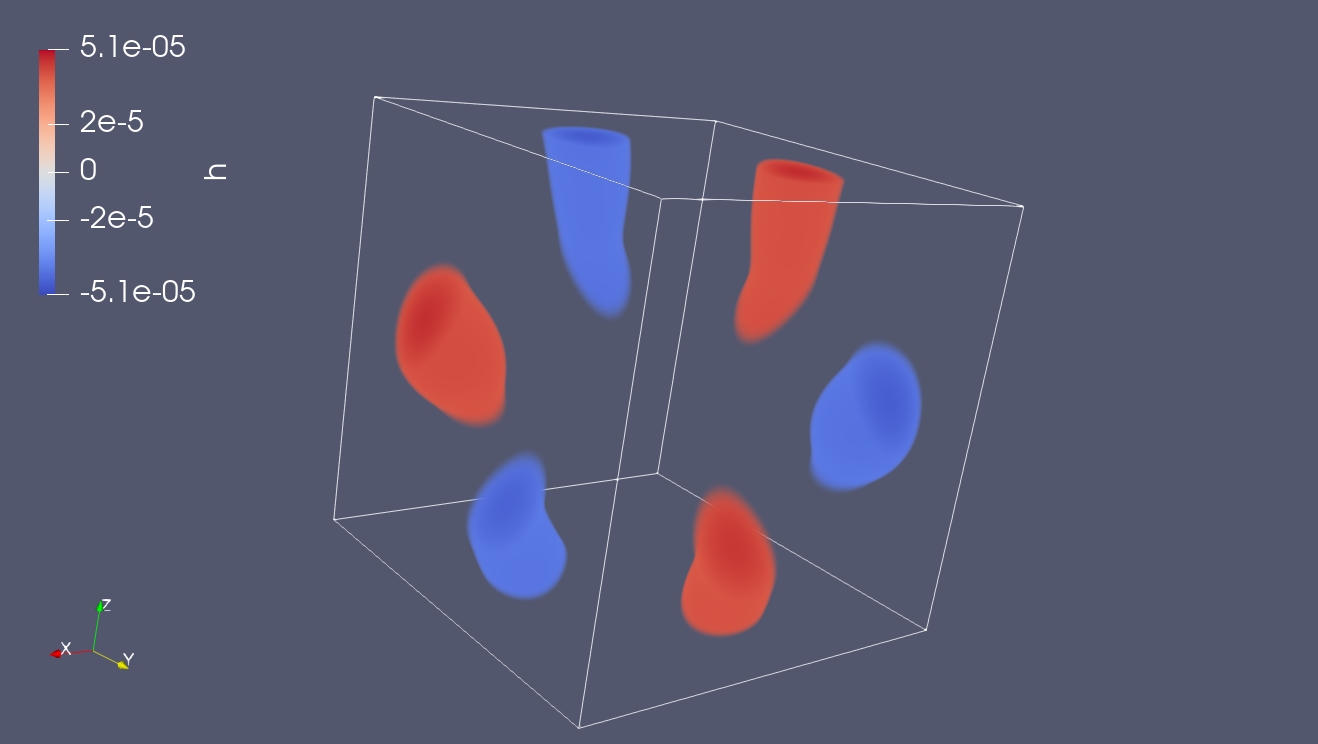}}
}
    
\caption{[Long time window, $T = 75$] Isosurfaces of the components
  (a) $\omega_1$, (b) $\omega_2$ and (c) $\omega_3$ of the vorticity
  field and {of (d) the helicity density $h(\x,0)$} corresponding
  to the optimal initial condition $\teta_{75}^{1024}$.}
    \label{fig:vort:xyz:3D:0:1024}
\end{figure}

Finally, we analyze the physical-space structure of the extreme flow,
beginning with the vorticity field of the optimal initial condition
$\widetilde{\bds \omega}_{75}^{1024} = [\omega_1, \omega_2, \omega_3]
= \bnabla \times \teta_{75}^{1024}$. In figure
\ref{fig:vort:xyz:3D:0:1024}, we show the three components of
$\widetilde{\bds \omega}_{75}^{1024}$ and observe that the optimal
initial condition has the form of three perpendicular pairs of
distorted anti-parallel vortex tubes.  If we define the component-wise
enstrophy as $\mathcal E_i := \frac{1}{2} \int_{\mbb T^3}
\omega_i^2(\bds x) \; d\bds x$, $i=1,2,3$, then we have $\mathcal E_1
= 6.34\times 10^{-5}\approx \mathcal E_2 = 6.35\times 10^{-5} <
\mathcal E_3 = 8.03\times 10^{-5}$, which is a signature of the
symmetry of the flow with respect to the plane $x_1 = x_2$.  {The
  helicity of the optimal flow defined as $\mathcal H(t) = \int_{\mbb
    T^3} h(\x,t)\; d \bds x$, where $h(\x,t) := \u(\x,t) \cdot
  \bomega(\x,t)$ is the helicity density, vanishes (it is an invariant
  of the evolution). The helicity density $h(\x,0)$ of the optimal
  initial condition $\teta_{75}^{1024}$ is shown in figure
  \ref{fig:vort:xyz:3D:0:1024}(d) revealing that the vanishing of the
  helicity $\mathcal{H}(t)$, $t \ge 0$, is a consequence of the
  aforementioned symmetry.}  In fact, this initial condition possesses
a similar physical-space structure as the one found in
\citet{KangYunProtas2020}, where the authors searched for
singularities in Navier-Stokes flows by maximizing the total enstrophy
at a prescribed final time. While, unlike here, no evidence for
singularity formation was revealed in that earlier study.

\begin{figure}
  \centering
\mbox{\subfigure[]{\includegraphics[width=0.48\textwidth]{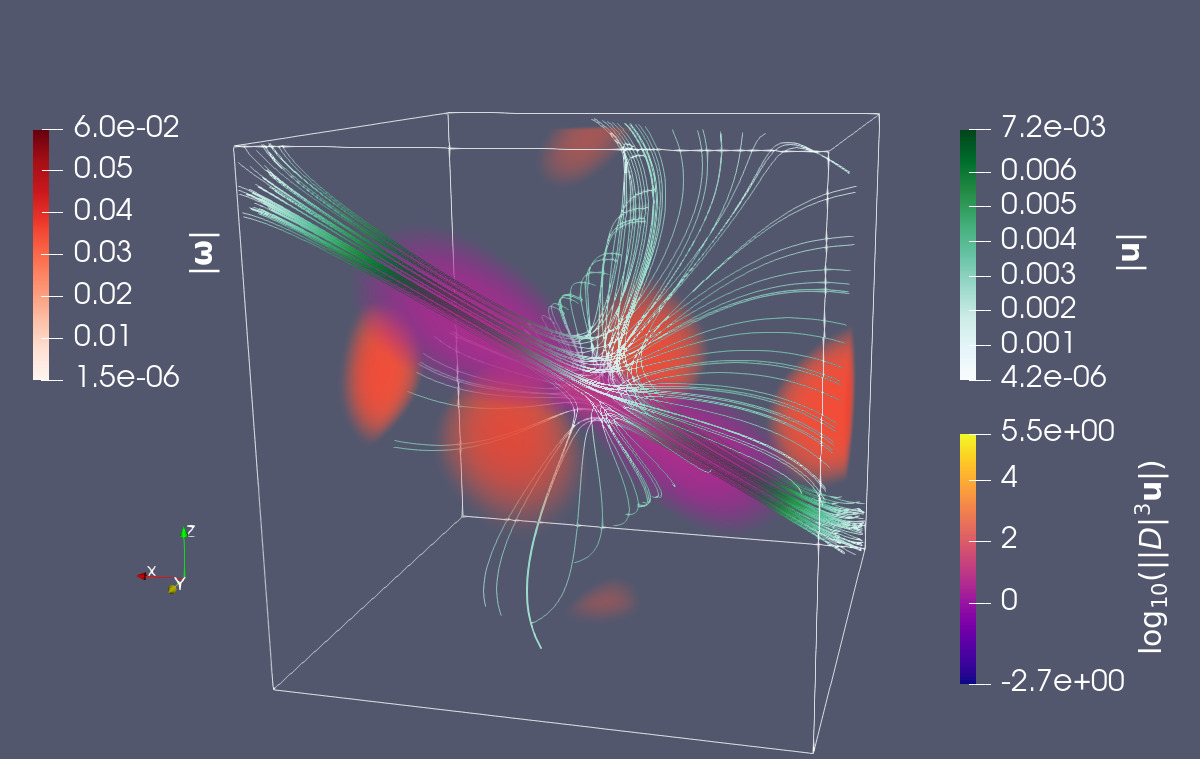}}
\hfill
\subfigure[]{\includegraphics[width=0.48\textwidth]{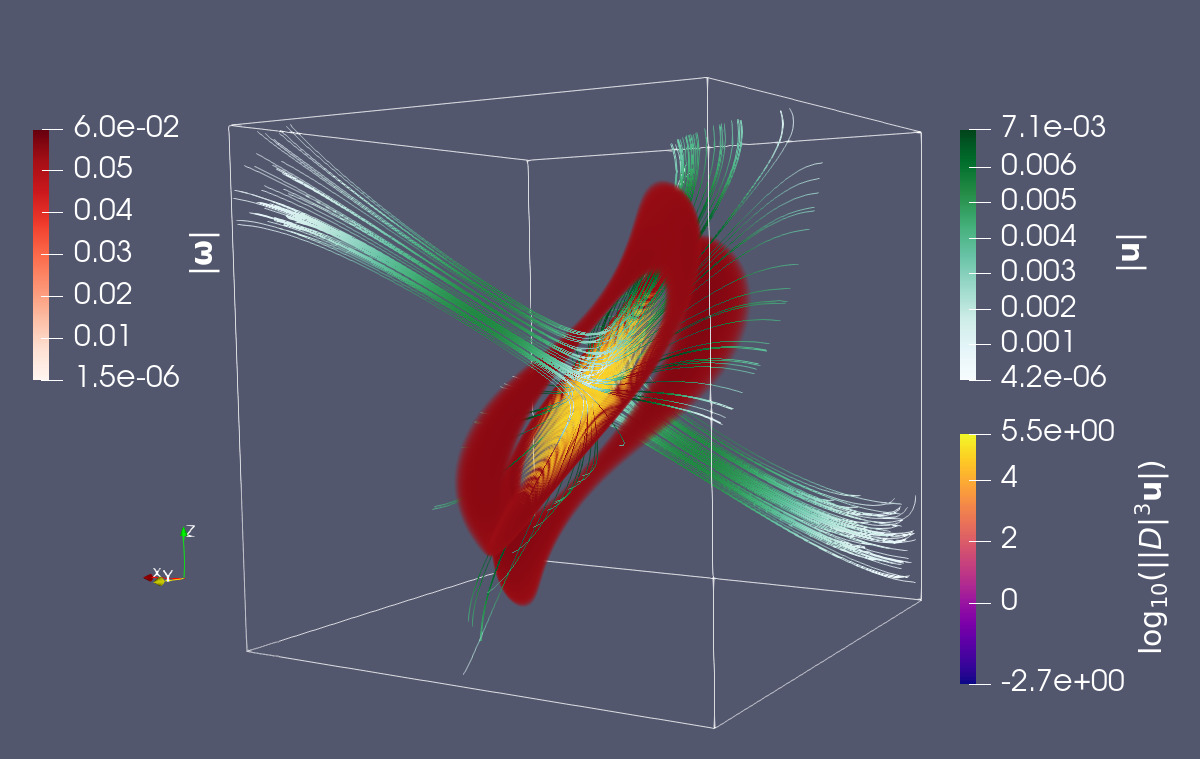}}}
\caption{[Long time window, $T = 75$] Isosurfaces of $|\bomega|$ and
  $\log_{10}(| |D|^3 \bds u|)$ in (a) the optimal initial condition
  $\teta_{75}^{1024}$ and (b) the corresponding terminal state
  $\u^{1024}\left(75; \teta_{75}^{1024}\right)$ together with selected
  streamlines. The same color ranges are used in both panels. An
  animated version of these figures (without streamlines) showing the
  time evolution for $t \in [0,75]$ is available as
  \href{https://youtu.be/G_nfJNq6W1A}{Movie 1}.  }
  \label{fig:vort:3D:75:1024}
\end{figure}

\begin{figure}
  \centering
\mbox{\subfigure[]{\includegraphics[width=0.48\textwidth]{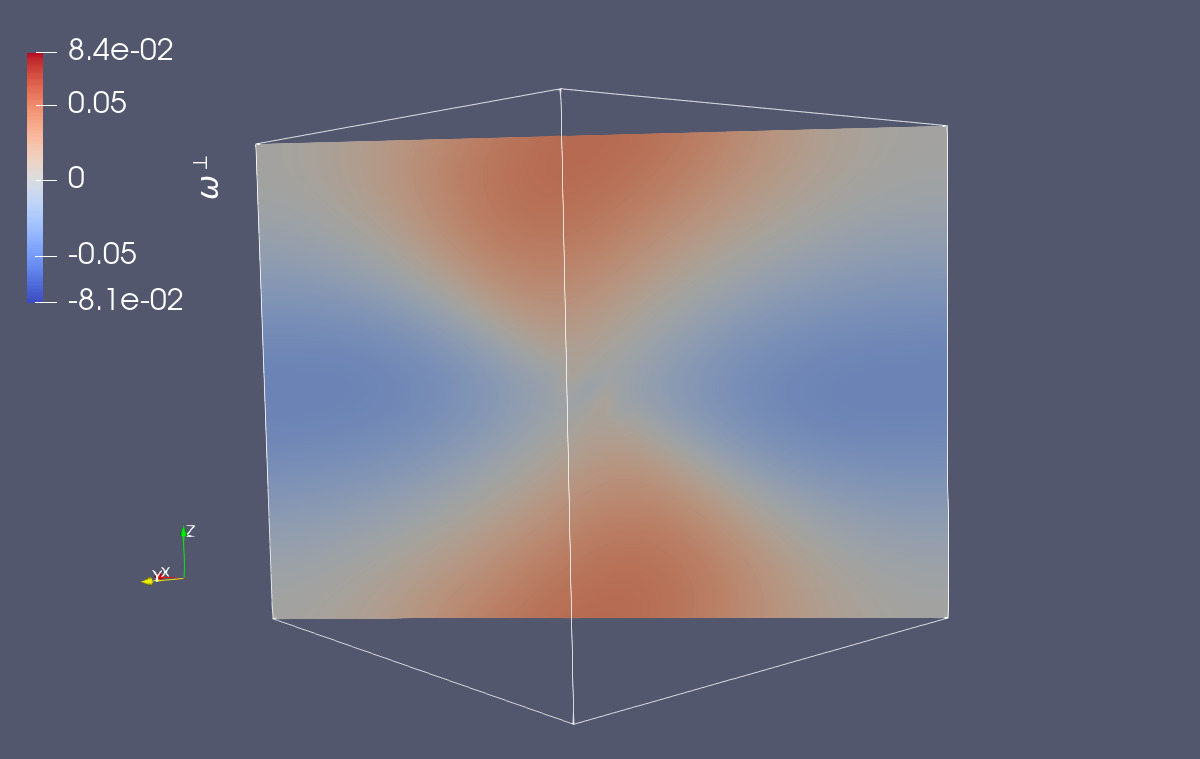}}
\hfill
\subfigure[]{\includegraphics[width=0.48\textwidth]{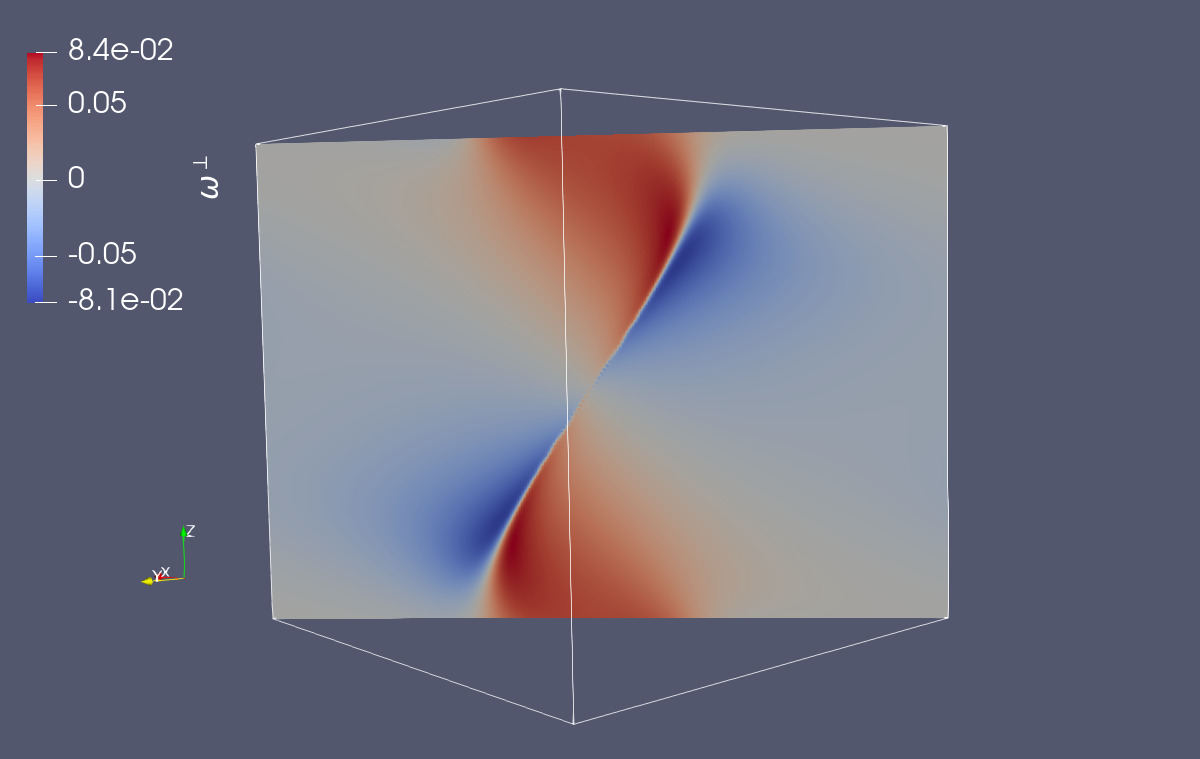}}}
\caption{[Long time window, $T = 75$] The vorticity component
  $\omega^\perp$ normal to the symmetry plane $x_1 = x_2$ in (a) the
  optimal initial condition $\teta_{75}^{1024}$ and (b) the
  corresponding terminal state $\u^{1024}\left(75;
    \teta_{75}^{1024}\right)$. The same color range is used in both
  panels. An animated version of these figures showing the time
  evolution for $t \in [0,75]$ is available as
  \href{https://youtu.be/6BcQ4LvcAE4}{Movie 2}.  }
  \label{fig:vort:proj:3D:75:1024}
\end{figure}

  Next, in figure \ref{fig:vort:3D:75:1024}(a), we analyze the optimal
  initial condition $\teta_{75}^{1024}$ in more detail with the
  corresponding terminal state $\u^{1024}\left(75;
    \teta_{75}^{1024}\right)$ shown in figure
  \ref{fig:vort:3D:75:1024}(b). In both figures we show the
  isosurfaces of $|\bomega|$ and $\log_{10}\left(\left| |D|^3 \bds
      u\right|\right)$ together with selected streamlines passing
  through a small neighborhood of the origin (see also
  \href{https://youtu.be/G\_nfJNq6W1A}{Movie 1} available in
  Supplementary Material).  We observe that throughout its evolution
  the optimal flow has the form of two jets colliding head-on with the
  vorticity concentrating into two strongly flattened vortex rings as
  time goes on. The region with large values of $\log_{10}\left(\left|
      |D|^3 \bds u\right|\right)$, which is the quantity that matters
  in our objective functional \eqref{eq:obj}, evolves into a flat disc
  located in the middle of the two rings.  The vorticity field at the
  final time $t = 75$ has three symmetry planes: $x_1 = x_2$, $x_1 =
  x_3$ and $x_2 = x_3$, in addition to discrete rotation symmetries
  with respect to the body diagonal passing through the center of the
  two rings.  Without loss of generality, we focus our discussion on
  the symmetry plane $x_1 = x_2$, and in figures
  \ref{fig:vort:proj:3D:75:1024}(a) and
  \ref{fig:vort:proj:3D:75:1024}(b) we visualize the vorticity
  component $\omega^\perp := \bds \omega \cdot \bds n$ normal to that
  plane ($\bds n = \left[-\frac{1}{\sqrt{2}}, \frac{1}{\sqrt 2},
    0\right]^T$ is the unit vector normal to the symmetry plane), at
  $t=0$ and $t=75$ (see also
  \href{https://youtu.be/6BcQ4LvcAE4}{Movie 2} available in
  Supplementary Material).
  The streamline pattern in figure~\ref{fig:vort:3D:75:1024}~(b)
  indicates that as $t$ approaches 75, the flow in the two jets
  colliding near the origin sharply transitions towards a radial
  outflow through the gap between the two vortex rings. At the same
  time, this gap becomes very narrow, which results in a sharp
  transition between the regions of the symmetry plane characterized
  by opposite signs of the normal vorticity $\omega^\perp$, cf.~figure
  \ref{fig:vort:proj:3D:75:1024}(b). This appears to be the mechanism
  responsible for the possible singularity formation in the extreme
  flow considered here.

%

\section{Discussion and conclusions} 
\label{sec:final}

In this paper, we study the problem concerning the
  possibility of spontaneous formation of singularities in solutions
of the 3D Euler equations \eqref{eq:Euler} on a periodic domain.
Based on the local well-posedness results by \citet{Kato1972} stated
in Theorem \ref{thm:Hm}, we formulate a PDE-constrained optimization
problem to search for initial conditions $\bds \eta$ with unit $\dot
H^3$ {seminorm} such that the corresponding optimal solution
  achieves a maximum {$\dot H^3$ seminorm} at a prescribed time $T$.  Since we
focus on smooth (real-analytic) initial data, the optimization
problem is formulated in a suitable Gevrey space \eqref{eq:G}.  It is
then solved in the ``optimize-then-discretize'' setting using a
state-of-the-art Riemannian conjugate gradient method \eqref{eq:RCG}
where the search direction at every iteration depends on the gradient
of the objective functional at the current iteration as well as on the
previous search direction; the former is obtained by solving
the adjoint system~\eqref{eq:adjoint} backward in time.  The required
regularity of the gradient is ensured by the use of the Riesz
representation theorem, cf.~\eqref{eq:dif:grad}. An analogous approach
has been successfully used to solve PDE-constrained optimization
problems formulated to elucidate extreme behaviors in 1D Burgers
\citep{ap11a} and 3D Navier-Stokes flows
\citep{KangYunProtas2020,KangProtas2021}.

Problem \ref{pb:H3} is non-convex and we have found evidence for the
presence of nonunique local maximizers corresponding to different
initial guesses listed in \S\; \ref{sec:IG}. However, iterations
performed with most initial guesses were found to converge to the same
(up to rotation and translation) local maximizer, with the
exception of the initial guess $\bds \eta_\Kerr$, which was designed
to produce significant growth of various regularity indicators at
times much longer than the intervals $[0, T]$ considered in
{this study.}

We adopt an indirect approach to determine whether a
  singularity may occur within the time interval $[0, T]$ based on
  sequential refinements of the numerical resolution.  Specifically,
  we solve Problem \ref{pb:H3} with $T = 25$ and $T = 75$ using
spatial resolutions increasing from $128^3$ to $1024^3$. For
$T = 25$, we observe that the maximum attained values of $\left|\left|
  \u^N\left(75;\teta_{75}^N\right) \right|\right|_{\dot{H}^3}$ converge to a finite
limit for the optimal solutions obtained with
increasing resolutions, which indicates that the Euler
system \eqref{eq:Euler} is well-posed on $[0, 25]$. However, for $T =
75$, the maximum values of $\left|\left|
  \u^N\left(75;\teta_{75}^N\right) \right|\right|_{\dot{H}^3}$ attained in the optimal
solutions diverge upon resolution refinement, which suggests a
possible formation of singularity for some $t \in [0, 75]$.
As shown in figure \ref{fig:u:H3:rate}, the growth of the
  norm $\| \u(t) \|_{\dot{H}^3}$ in the flows corresponding to the
optimal initial condition obtained for $T = 75$ using the
highest numerical resolution proceeds at a rate consistent with a
finite-time blow-up as long as the solution remains
well-resolved, although this rate slows down with time. With the
limited numerical resolution we could use, it is impossible to
conclude whether this depletion of growth rate would eventually
prevent a singularity from appearing in a finite time. We add that the
largest resolution we used ($1024^3$) is smaller than the largest
resolutions available to-date in simulations of the Navier-Stokes and
the Euler systems. However, we emphasize that in the present study we
solve a family of optimization problems which are much more costly to
solve than simply computing the evolution of the solution on the same
time interval. In fact, solution of Problem \ref{pb:H3} typically
requires $\mathcal{O}(100)$ solutions of the Euler system
\eqref{eq:Euler} and $\mathcal{O}(10)$ solutions of the adjoint system
\eqref{eq:adjoint}.

We add that under-resolved computations of Euler flows lead
  to ``thermalization'' of solutions, a process where after a
  sufficiently long time the kinetic energy is equidistributed among
  the finite number of Fourier modes used in the computation.
  Thermalization is related to the appearance of the so-called
  ``tygers'' and inevitably occurs when using numerical methods based
  on Galerkin truncation to compute the time evolution of inviscid
  systems, such as 1D Burgers equation \citep{Rampf2022} and 3D
 incompressible Euler equations \citep{Murugan2022}.

Figures \ref{fig:norms:25} and \ref{fig:H3:max:compare:75}
  show that, as compared to flows obtained with different initial
  guesses, the optimal flows demonstrate a more significant growth in
  both $\|\u(t; \teta) \|_{\dot{H}^3}$ and $\| \bomega(t;
\teta)\|_{L^\infty}$ throughout the optimization time windows $[0, T]$
with $T = 25$ and $T = 75$.  We further notice that the growth of the
norm $\| \bomega(t) \|_{L^\infty}$ in the extreme flows obtained by
solving Problem \ref{pb:H3} with $T = 75$ is weaker than the growth of
the norm $\| \u(t) \|_{\dot{H}^3}$ in the same flow,
cf.~figures \ref{fig:H3:max:compare:75}(a) and
\ref{fig:H3:max:compare:75}(b).  However, estimate
\eqref{eq:BKM:necessary} shows that an unbounded growth of the latter
norm implies such a growth for the former norm as well, and
the doubly-exponential structure of the upper bound in this
  estimate explains why in a blow-up scenario the growth of $\|
\bomega(t) \|_{L^\infty}$ may be weaker than that of $\| \u(t)
\|_{\dot{H}^3}$.  As regards the behavior of the width $\delta(t)$ of
the analyticity strip in these flows, we note that the decay rate of
this quantity accelerates and becomes faster than exponential around
the time when computations become under-resolved.

In regard to the structure of the extreme flows in the physical space,
it is interesting to see in figure \ref{fig:vort:3D:75:1024}(b) that
some key features of these flows are close to being axisymmetric, to
the extent that this may be possible for a flow evolving on a 3D
periodic domain.  More specifically, the region where the largest
values of $\left| |D|^3 \bds u\right|$ are concentrated has the form
of a circular disc embedded between the two colliding vortex rings,
which becomes more irregular further away from the axis of symmetry.
The center of this disk, located on the symmetry plane of the extreme
flows, is a stagnation point; in this sense, the extreme flows are
similar to the infinite-energy solutions of the Euler system
constructed by \citet{Gibbon1999}, which are known to exhibit
finite-time singularities \citep{MulungyeLucasBustamante2016}.  We
emphasize that the extreme flows we found here are fairly robust in
the sense that they have been found consistently using several
different initial guesses in \eqref{eq:RCG}, albeit not with all
initial guesses we used. {We add that a head-on collision of
  two vortex rings has served as a classical paradigm in recent
  experimental and numerical studies of turbulence generation
  \citep{mckeown2018cascade, lim1992instability,
    mckeown2020turbulence}. At high Reynolds numbers, these two vortex
  rings break down into smaller antiparallel secondary and tertiary
  filaments, eventually forming a turbulent cloud due to the elliptic
  instability \citep{mckeown2020turbulence}. }

As concerns future work, the properties of the optimal flows found in
the present study suggest looking for extreme behaviors among initial
conditions constrained by some symmetries, such as axial and/or
reflection symmetries, as was pursued by \citet{lh14a,
  lh14b,Hou:22:Euler}. In addition, given that spectral methods have a
limited ability to resolve localized small-scale features of
potentially singular flows, further progress will likely require the
use of adaptive discretization techniques. 

 

\section*{Acknowledgements}

The authors wish to express thanks to Marc Brachet, Miguel Bustamante,
John Gibbon, Robert Kerr, Adam Larios and Thomas Y.~Hou for
enlightening and enjoyable discussions.The authors also acknowledge
the support through an NSERC (Canada) Discovery Grant.  Computational
resources were provided by Digital Research Alliance of Canada under
its Resource Allocation Competition.



\FloatBarrier


\end{document}